\documentclass{aa}  

\usepackage[dvipsnames]{xcolor}
\usepackage[breaklinks, colorlinks, citecolor=blue]{hyperref}
\usepackage{natbib}
\usepackage{graphicx}
\usepackage{txfonts}
\usepackage{hyperref}
\usepackage{ulem}  

\begin{document} 
  \title{Dissecting a miniature universe: \\
  A multi-wavelength view of galaxy quenching in the Shapley supercluster}
  \subtitle{}

  \author{N. Aghanim\inst{1}\thanks{nabila.aghanim@ias.u-psud.fr}
     \and
         T. Tuominen\inst{1}
     \and
         V. Bonjean\inst{2,3}
     \and
         C. Gouin\inst{1}     
    \and
         T. Bonnaire\inst{4} 
     \and
        M. Einasto\inst{5}
          }

  \institute{Université Paris-Saclay, CNRS, Institut d’Astrophysique Spatiale, 91405 Orsay, France
         \and
             Littoral, Environnement et Sociétés, Université de La Rochelle, and CNRS (UMR7266), La Rochelle, France
        \and
Instituto de Astrof\'{i}sica de Canarias, E-38205 Tenerife, and University of La Laguna, E-38206 Tenerife, Spain
      \and
     Laboratoire de Physique de l’École normale supérieure, ENS, Université PSL, CNRS, Sorbonne Université, Université Paris Cité, F-75005 Paris, France
     \and
     Tartu Observatory, Tartu University, Observatooriumi 1, 61602 Tõravere, Tartumaa, Estonia
     }

  \date{}

  \abstract{ 
Multiple-cluster systems, i.e. superclusters, contain large numbers of galaxies assembled in clusters inter-connected by multi-scale filamentary networks. As such, superclusters are a smaller version of the cosmic web and can hence be considered as miniature universes. In addition to the galaxies, superclusters also contain gas, hot in the clusters and warmer in the filaments. Therefore, they are ideal laboratories to study the interplay between the galaxies and the gas. In this context, the Shapley supercluster (SSC) stands out since it hosts the highest number of galaxies in the local universe with clusters interconnected by filaments. In addition, it is detected in both X-rays and via the thermal Sunyaev-Zel'dovich (tSZ) effect, making it ideal for a multi-wavelength study of the gas and galaxies. Applying for the first time a filament-finder based on graphs, T-REx, on a spectroscopic galaxy catalogue, we uncovered the 3D filamentary network in and around SSC. Simultaneously, we used a large sample of photometric galaxies with information on their star formation rates (SFR) in order to investigate the quenching of star formation in the SSC environments which we define as a function of the gas distribution in the \textit{Planck} tSZ map and the ROSAT X-ray map.\\ 
With T-REx, we confirm filaments already observed in the distribution of galaxies of the SSC, and detect new ones. We observe the quenching of star formation as a function of the gas contained in the SSC. We show a general trend of decreasing SFR where the tSZ and X-ray signals are the highest, within the high density environments of the SSC. Within these regions, we also observe a rapid decline of the number of star-forming galaxies, coinciding with an increasing number of transitioning and passive galaxies. Within the SSC filaments, the fraction of passive galaxies is larger than outside filaments, irrespective of the gas pressure. Our results suggest the zone of influence of the SSC, in which galaxies are pre-processed and quenched, is well defined by the tSZ signal that combines the density and temperature of the environments.

} 


\maketitle


\section{Introduction}

The cosmic web is an entangled structure made of nodes (galaxy groups and clusters) at the intersection of filaments surrounded by walls and separated by voids. In the cosmic web, galaxy
superclusters are defined as the high-density regions in the distribution of galaxies or galaxy groups/clusters \citep[e.g.,][]{maret-cat1997,liivamagi2012,chon2013,chow2014,superclusSDSS2023,tartu2023}. Superclusters occupy only 1\% of the volume of the Universe; however they host around 15\% of galaxies including all very rich and luminous galaxy clusters which reside in superclusters or in their high-density cores \citep[e.g.,][]{maret-2022,maret-2023arxiv}. Furthermore, numerical simulations show that the progenitors of the present-day galaxy superclusters are already seen in the very early Universe \citep{2022ApJ...937...15P,maret-2023}.
Several studies have shown that galaxy superclusters or their high-density cores are the largest objects which are either collapsing now or which will collapse in the
future \citep[e.g.,][]{dunner2006,ursamajor-grav2007,luparello2011,chon2015,maret-wall2016,maret2021,maret-2022}. This property has further been used to define superclusters as in \citet{luparello2011} and \citet{chon2015}. In their study, \citet{maret-2022} compared different definitions of superclusters based on their sizes
and masses. They showed that the largest collapsing cores are associated with the richest superclusters, such as the Shapley and the Corona Borealis supercluster in the local Universe, or the BOSS Great Wall at redshift $z = 0.47$. These cores have sizes up to approximately $10 - 12$ h$^{-1}$~Mpc and masses up to $10^{16} M_{\mathrm{sun}}$.

Superclusters embed galaxies, groups, and clusters connected by galaxy filaments, which may extend to their surrounding cosmic web. Supercluster environment is known to
impact the evolution of galaxies within them. This was shown in seminal studies \cite[e.g.,][]{maret1987} and in detailed analyses of a few nearby superclusters \citep[e.g.,][]{luparello2013,pearson2014,lietzen2016,lacerne2016,maret2142-2018,maret-2020,maret2021,castignani2022,pekkaSC2022}. 
They conclude that galaxies in groups within superclusters have older stellar populations,
suggesting that groups in superclusters formed differently from groups in other environments \cite[see e.g.,][ for a recent study]{maret-2023} with their galaxies impacted  via a stage of pre-processing before they fall in the clusters. 
In addition to their galaxy content, simulations \citep[e.g.,][]{Flores-simu-gasSC2009} show that superclusters contain large quantities of hot ionised gas which should be observed via the thermal Sunyaev-Zel'dovich (tSZ) signal \citep{Sunyaev1970}. As a matter of fact, evidence for hot ionised gas in superclusters was found either in individual cases \citep[e.g.,][]{tSZ-corona2006}, for a few selected objects \cite[e.g.,][]{pekkaSC2022} or via the stacking of hundreds of superclusters \citep{hidekiSC2019} or directional stacking \citep{lokken-orientedStack2022}. This gas was also observed in a few cases in the X-ray domain \citep[e.g.,][]{shapleyX1999,reiprich2021,erosita-33-2022}. The interplay between gas and galaxies in clusters is one of the main drivers of galaxy evolution and as such it has been extensively studied. This interplay intervenes at different redshifts and masses, it is quite complex, and it takes very diverse forms  \citep[such as ram pressure stripping, evaporation, etc., see e.g., ][for early studies]{gunn1972,cowie1977,larson1980,nulsen1982}. In addition to the impact of the local overdensity, those physical processes actually shape the galaxy properties and their star formation activity. 

\medskip
In the present study, we investigate the relation between the hot gas in a supercluster and the galaxy properties. To do so, we determine the supercluster's structure and examine how the star-formation rate of the galaxy populations evolves with the Compton $y$ parameter tracing the hot gas distribution. \\
For our analysis, we focus on the extended cosmic-web structure that contains the largest concentration of galaxies and rich Abell clusters \citep[34 in the catalogue of ][]{maret-cat1997} in the local Universe, namely the Shapley supercluster \citep[SSC,][]{1930BHarO.874....9S}. 
This miniature Universe extends over $\approx 260$~Mpc$^2$ and covers a redshift range from 0.033 to 0.06 (with a mean redshift $z = 0.048$). Its core region, with the right ascension ranging between R.A. = 198 and 204
degrees and declination ranging between -34 and -29
degrees, includes 11 clusters and groups (A3552, A3554, A3556, A3558, A3559, A3560, A3562, AS0724, AS0726, SC1327-312, SC1329-313) with masses in the range $M_{500} \approx 1 - 6 \times 10^{14}\,M_{\mathrm{sun}}$. Among these, A3558 and A3528 are merging X-ray clusters which form two complex structures in the centre and in the outskirts of SSC \citep{bardelli-2004}. The high-density core of the Shapley supercluster is probably the largest among
the collapsing  cores of superclusters in the local Universe \citep{reisenegger2000} with a mass and radius of  $M \approx 1.3\times 10^{16}\,M_{\mathrm{sun}}$ and radius $12.4\,h^{-1}\,$Mpc. Our choice of the SSC as a case study is not only due to its exceptional nature; it is also triggered by the wealth of data available for a multi-wavelength analysis. Indeed, the SSC was lengthly observed and studied in the optical and nearIR \citep[e.g.,][]{quintana1995,quintana1997,mercurio2006,merluzzi2015}. It was observed in the X-rays \citep{shapleyX1999,merluzzi2016} and it is seen in the ROSAT survey. Finally, Shapley's hot ionised gas is clearly observed via tSZ in the \textit{Planck} data \citep{plck-shapley2014,HidekitSZ}.\\
The paper is organised as follows. In Sect. \ref{sec:data}, we describe the photometric and spectroscopic data used to study the structure of SSC and the galaxy properties, together with the data tracing the hot ionised gas. In Sect. \ref{anal}, we present the results we obtained from our different analyses. We start by addressing the SSC structure as revealed by the detection of filaments with a state-of-the-art technique. We continue with the distribution of the galaxy populations in the SSC field. Finally, we exhibit the relation between the star-formation rate in galaxies and the Compton $y$ parameter and investigate its modification in the environment defined by the SSC filamentary pattern. We discuss our results in Sect. \ref{discussion} and conclude in Sect. \ref{conclusion}. In the paper, we used $H_0=70$km/s/Mpc.

\begin{figure*}
\centering
\includegraphics[width=\hsize]{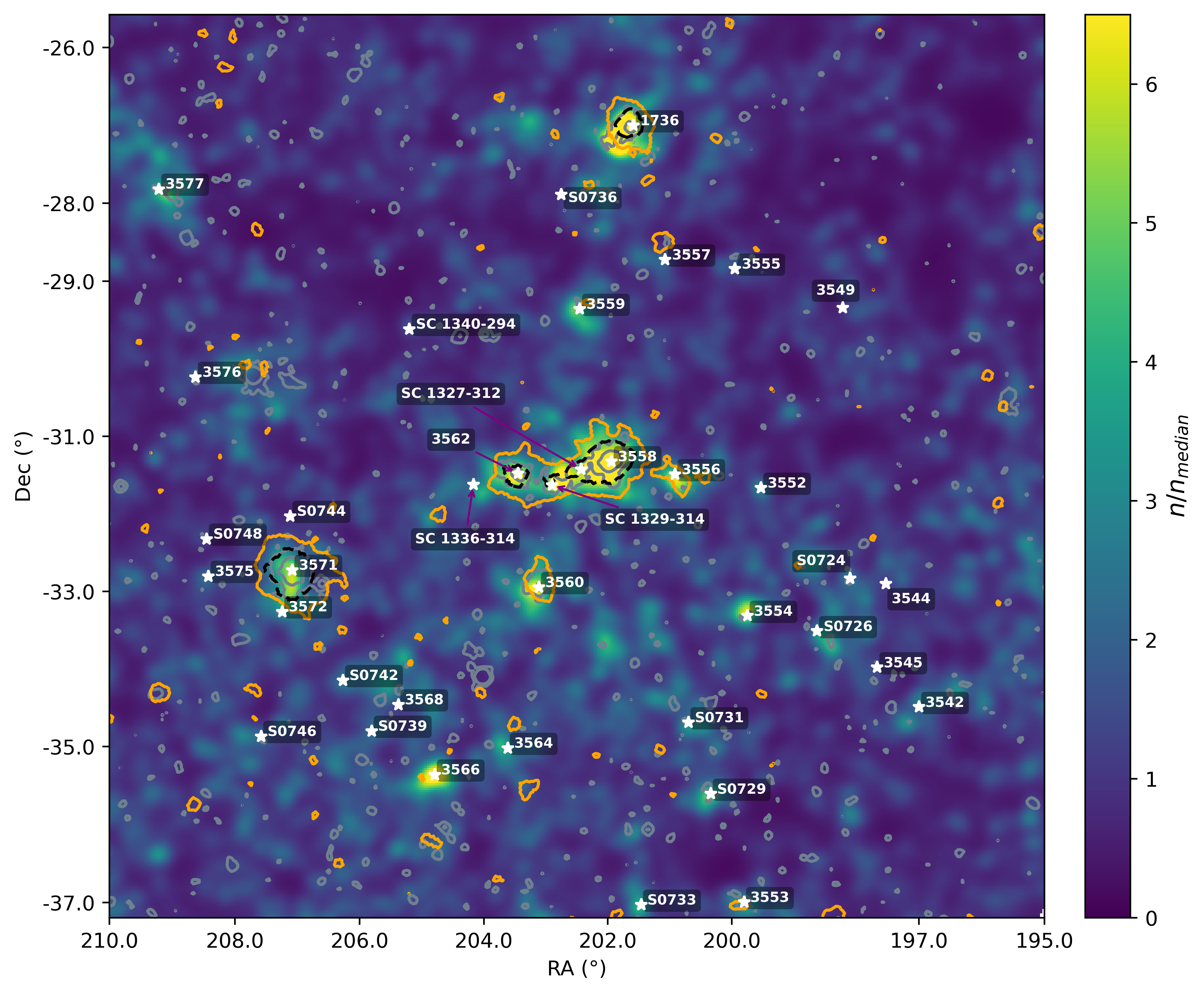}
\caption{Image, centred on Shapley Super Cluster, within a redshift range of $0.03<z<0.083$. White stars indicate known clusters from \citep{Proust2006A&Acatalogue}. The colour scale represents the galaxy number density, normalised by the median density in each pixel. The outer orange solid contours and inner black dashed contours correspond to Compton $y$ parameters values of $4.4\times 10^{-6}$ and $1.1\times 10^{-5}$, respectively, in the tSZ reconstructed map from the \textit{Planck} all-sky survey. The outer and inner grey contours correspond to X-ray signal of 2000 counts$\,$s$^{-1}\,$arcmin$^{-2}$ and $10^{4}$ counts$\,$s$^{-1}\,$arcmin$^{-2}$, respectively, in the ROSAT all-sky survey map.}
\label{FigFull_map}
\end{figure*}

\section{Data}\label{sec:data}
\subsection{Photometric galaxy catologue}\label{photGal}

The photometric galaxy catalogue used in this study was specifically designed to the analysis of the Shapley supercluster, namely covering the large sky area and including key galaxy properties such as star-formation rate and stellar masses. It is an optimised version of the value added catalogue constructed using machine learning following \cite{bonjean2019} from the WISExSCOS photometric redshift catalogue \citep{bilicki2016}, for the faint galaxies, combined with 2MASSxSCOSxWISE for the bright galaxies. The approach followed by \cite{bonjean2019}, based on the training of a Random Forest (RF) algorithm, shows that estimated star-formation rate (SFR) and stellar masses ($\mathrm{M}_\star$) of both passive and star-forming galaxies can be derived accurately, i.e. with reduced scatter and are unbiased with respect to redshift in the range of the training sample \cite[see][for comparison with the literature]{bonjean2019}.

\begin{figure*}
   \centering
   \includegraphics[width=0.45\textwidth]{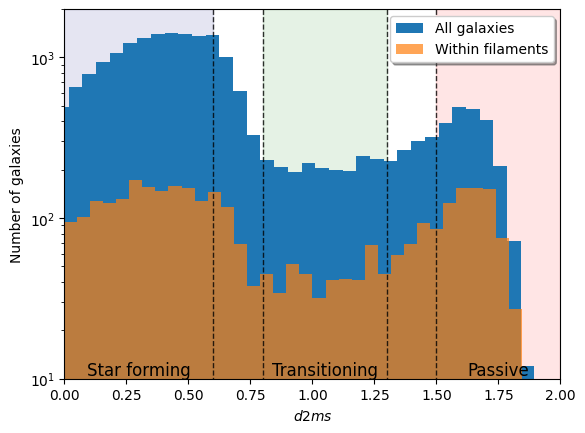}
   \includegraphics[width=0.45\textwidth]{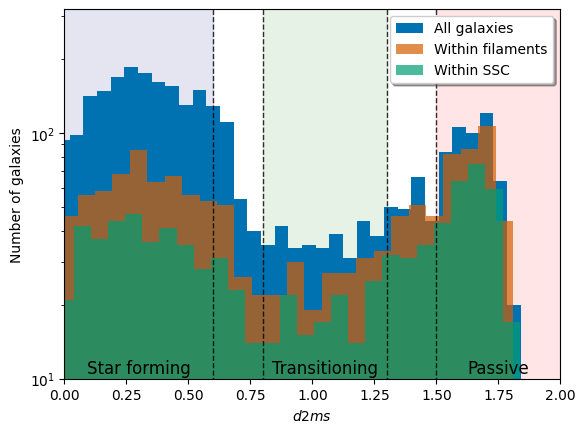}
   \caption{\textit{Left panel:} Number of photometric galaxies in different populations, as defined by their distance-to-main-sequence, d2ms, for the full Shapley supercluster area shown in Fig. \ref{FigFull_map}. Blue bars show all the galaxies while orange bars indicate the number of galaxies within filaments. \textit{Right panel:} Same as left for the zoom-in region discussed in Sect. \ref{sec:SSCenv}, with the number of galaxies within the SSC core region (i.e. with a Compton-$y$ values $y > 2.2 \times 10^{-6}$, see Fig \ref{FigPopulations}) in green.   
   }
    \label{Fig:d2ms_histogram}%
    \end{figure*}

\begin{figure*}
   \centering
   \includegraphics[width=0.45\textwidth]{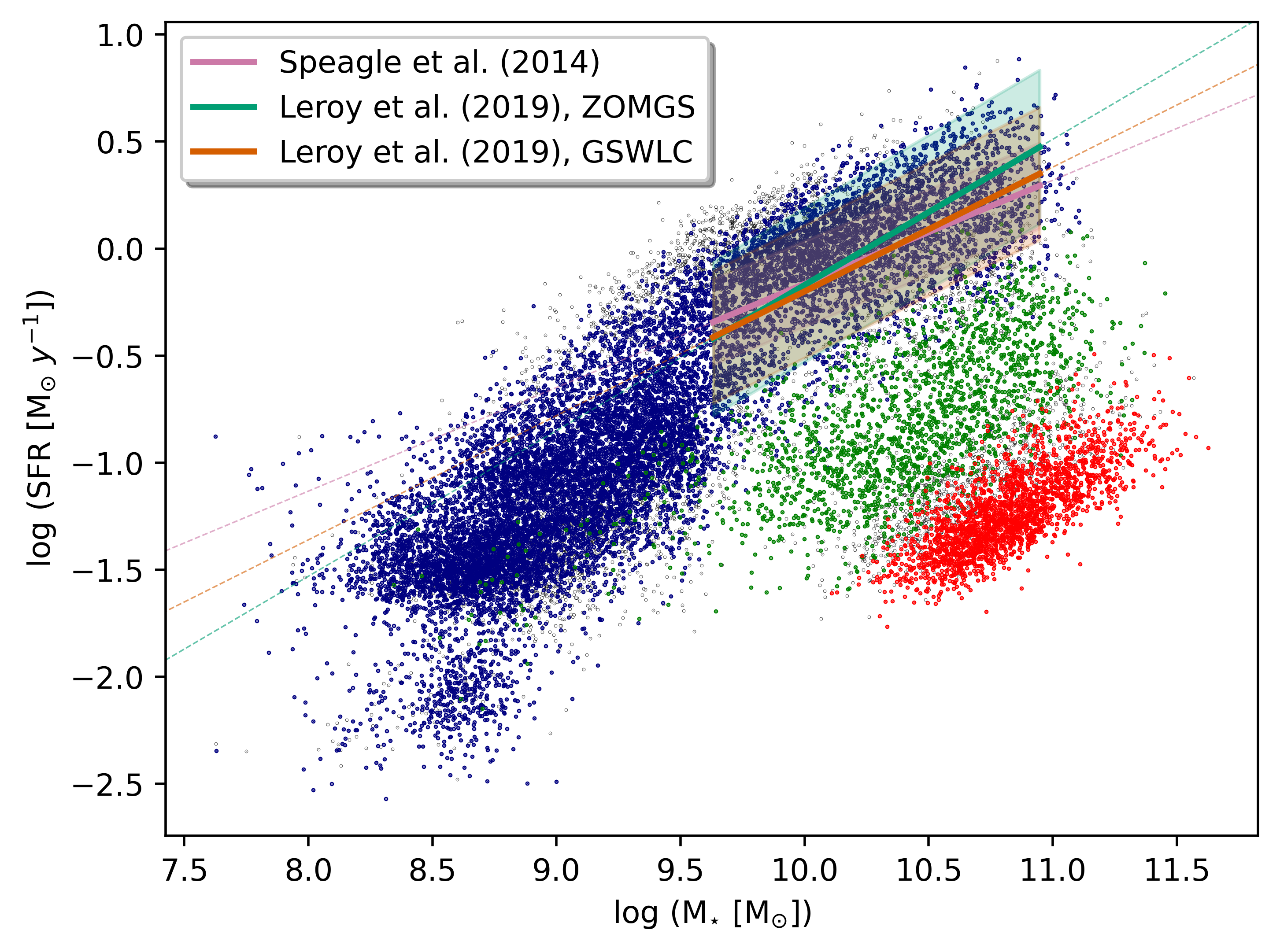}
   \includegraphics[width=0.45\textwidth]{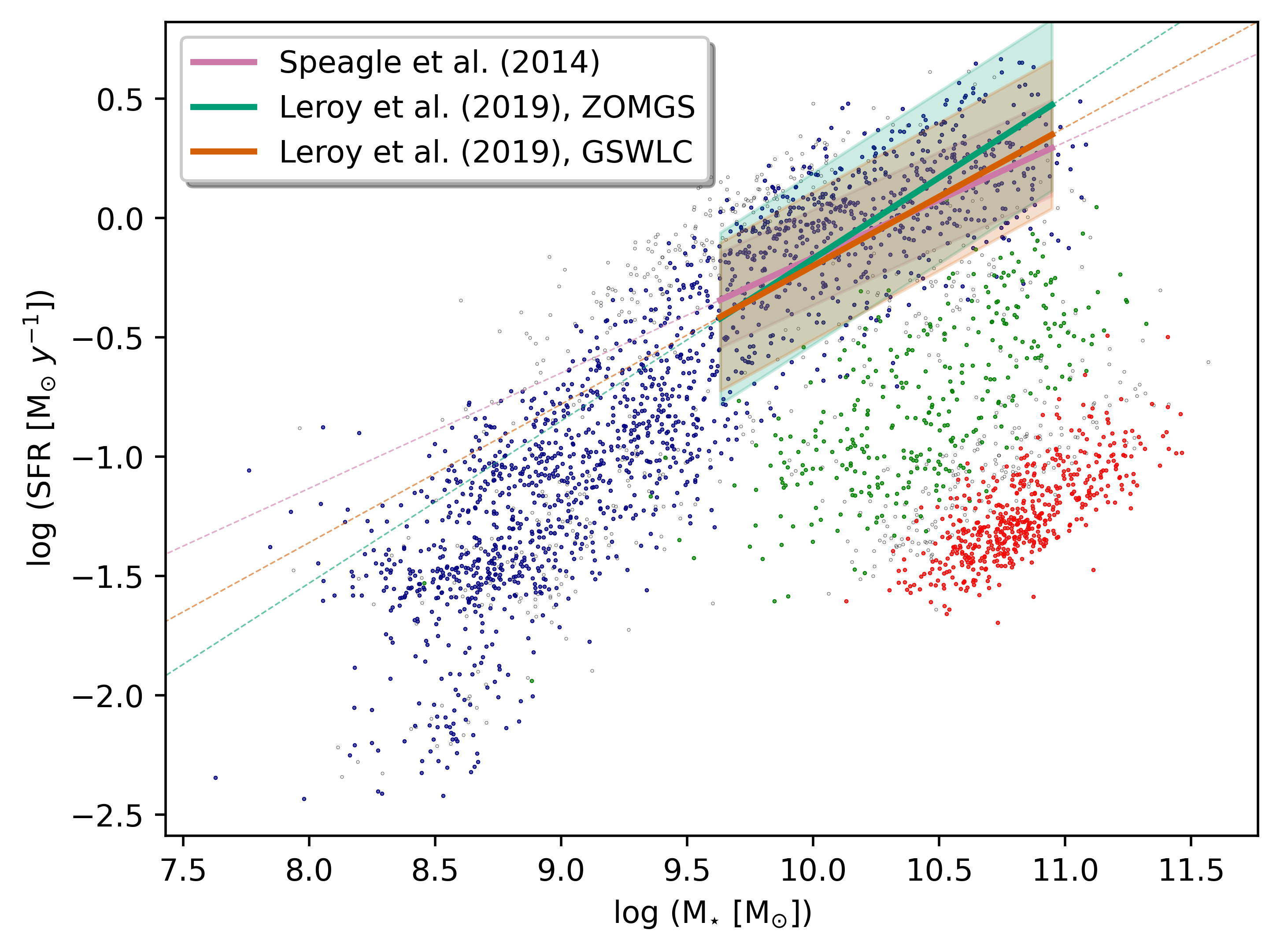}
   \caption{\textit{Left panel:} SFR - M$_{\star}$ diagram of the galaxies from the value-added photometric catalogue within the SSC field (Fig. \ref{FigFull_map}). Blue dots correspond to star-forming galaxies, green are transitioning while red dots are passive galaxies, defined by their d2ms (see Fig. \ref{Fig:d2ms_histogram}). Gray dots are galaxies not clearly defined within any of the populations (between shaded areas in Fig. \ref{Fig:d2ms_histogram}). The coloured lines correspond to scaling relations computed by \cite[][violet]{speagle2014} and \cite{leroy2019} from ZOMGS (turquoise) and GSWLC (orange) galaxy catalogues, with reported scatters of 0.2, 0.36 and 0.31 dex, respectively (shaded regions). The solid lines and shaded areas correspond to the mass range used for computing the scaling relation ($9.5 < \log (\mathrm{M}_\star) < 11$) \textit{Right panel:} Same as left for the zoom-in region discussed in Sect. \ref{sec:SSCenv}.   
   }
    \label{Fig:mass_SFR}%
    \end{figure*}

By construction, the accuracy of the RF-based estimates of SFR and $\mathrm{M}_\star$ depends on the learned-model, i.e. on the training sample and more specifically its redshift range. Therefore, we have defined a redshift range encompassing the Shapley supercluster ($z < 0.15$) and applied the approach of \cite{bonjean2019}. Using the redshift estimates from the two catalogues (WISExSCOS and 2MASSxSCOSxWISE), the SCOS magnitudes in the R and B bands, and the WISE magnitudes in the bands W1 and W2, we estimated the star-formation rate and stellar masses for the 83,700 galaxies in the range $z < 0.15$. Following, \cite{bonjean2019}, we have used the distance to the main sequence on the SFR-$\mathrm{M}_\star$ diagram (d2ms, hereafter), which estimates the star-formation activity of galaxies, as a criterion to define three baseline populations of galaxies, namely star-forming (0.0 < d2ms < 0.6), transitioning (0.8 < d2ms < 1.3), and passive (1.5 < d2ms < 2.0). The ranges were chosen  such that each population is clearly distinct from the other two (as displayed in Fig. \ref{Fig:d2ms_histogram}).  In the redshift range $0.03 < z < 0.08$ including the Shapley supercluster complex 19195 galaxies are considered, among them 12542 are star-forming, 1907 are transitioning, and 2115 passive galaxies. The range of SFR and stellar masses of these populations are shown in Fig. \ref{Fig:mass_SFR}. We also show, as shaded areas covering the mass range use for their estimates, the local main sequence scaling relations derived by \cite{speagle2014}, in the range $9.7 < $log($\mathrm{M}_\star$)$ < 11.1$, and by \cite{leroy2019}, in the range $9.5 < $log($\mathrm{M}_\star$)$ < 11$. We see the good agreement between the different estimates in the local universe and those obtained by the Machine learning techniques used here, confirming the already noted agreement] between results in \cite{bonjean2019} and those in \cite{cluver2014} and \cite{wen2013} at higher redshift.

\subsection{Photometric galaxy maps}\label{photGalmap}
In the present study, we analyse the spatial distribution of the galaxies and their properties as compared to the gas distribution. Hence based on the added-value photometric galaxy catalogue described in Sect. \ref{photGal} and following \cite{bonjean2020}, we constructed full-sky maps of galaxy number density in the redshift range $0.03<z<0.083$ chosen so that galaxies belong to the Shapley supercluster, within the dispersion of the photometric redshifts $\sigma_z = 0.033$. Given this dispersion and the considered redshift range ($0.03<z<0.083$), it is not possible to define bins of redshifts; we hence considered a single redshift slice with the coordinates of each galaxy on the plane of sky given in the catalogue. We compute the galaxy number density on a continuous grid by applying the Delauney Tesselation Field Estimator \cite[DTFE,][]{dtfe2007} to the galaxy distribution.\footnote{The DTFE Python code developed by V. Bonjean can be found at \url{https://git.ias.u-psud.fr/ByoPiC/pydtfe}} We thus obtain a two-dimensional density field which we smooth, with a Gaussian smoothing kernel of radius $r \approx 0.04$ degrees, in order to reduce possible edge effects induced by the tesselation.

In addition to the number density map of the whole galaxy population, we compute in the same way the maps of the three different populations (star-forming, transitioning and passive galaxies). We also produce the associated SFR maps by computing the mean SFR within each pixel of the density maps. 

We show, in Fig. \ref{FigFull_map}, a $15\times15$ square degree area centred on Shapley where the colour map corresponds to the galaxy number density normalised by the median density. We display groups and clusters of galaxies detected in the SSC area in the X-rays, and listed in \cite{higuchi2020}, and in the optical from \cite{Proust2006A&Acatalogue}. They are shown as white stars. For each galaxy population (Fig. \ref{FigPopulations}), we display the galaxy number density normalised by the total galaxy number density. Thus, each map indicates the regions predominantly dominated by star-forming, transitioning or passive galaxies.

\subsection{Spectroscopic galaxies} \label{specGal}
\begin{figure}
   \centering
   \includegraphics[width=\hsize]{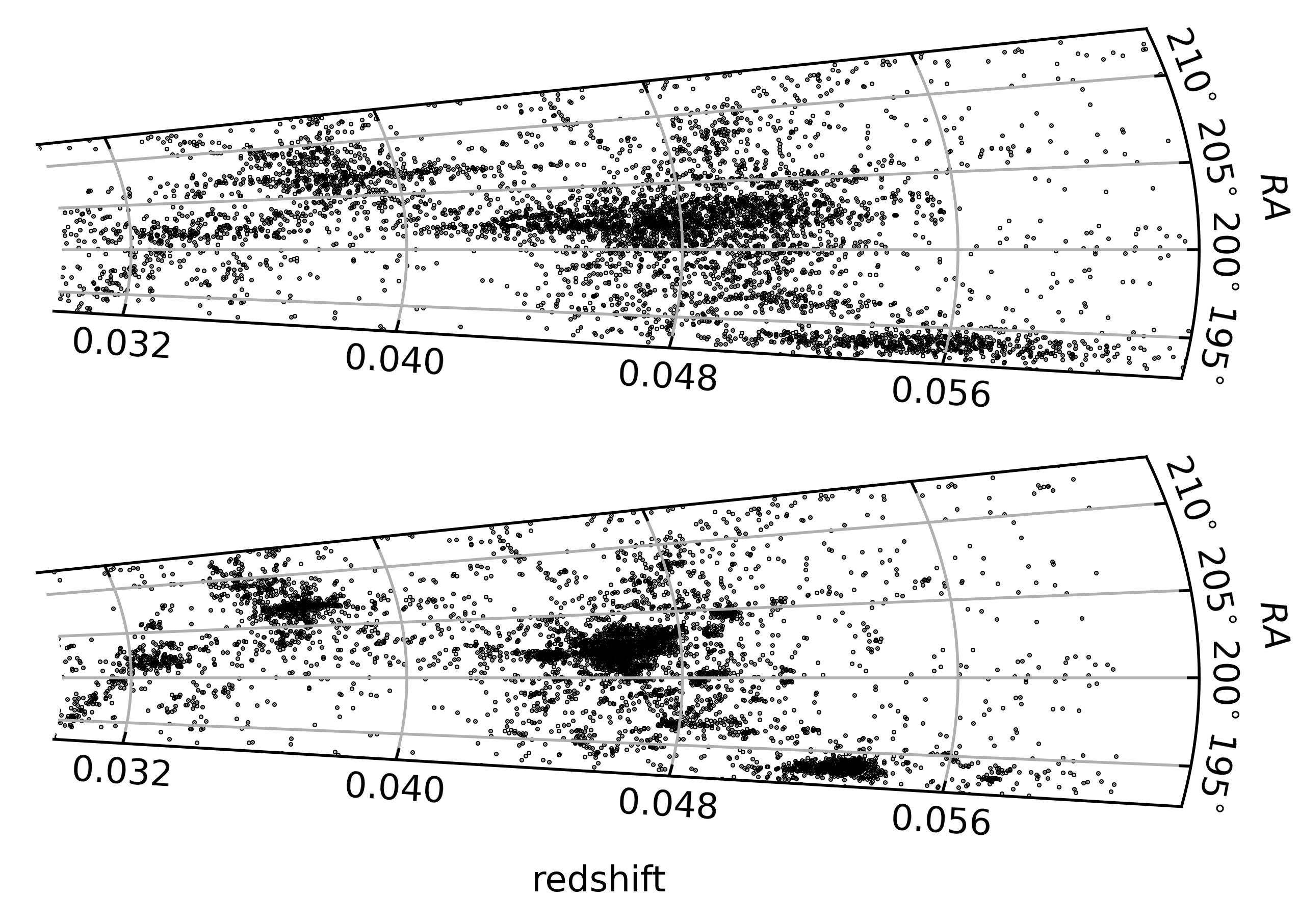}
      \caption{\textit{Upper panel}: Each black dot corresponds to a spectroscopic galaxies selected from \citet{quintana20} such that $R2<18$ mag and $9,000 <v< 18,900$ km/s. \textit{Lower panel}: Same galaxies after correcting for the FoG effect following Sect. \ref{specGal}.
              }
         \label{FigFoG_correction}
   \end{figure}

We have used the \textit{Shapley Supercluster velocity Database}. This is a publicly available and comprehensive catalogue of velocities in the area of about $300$ square degrees centred on SSC compiled by \cite{quintana20}. The dataset includes 18,146 velocity measurements with proper photometry and astrometry for 10,719 galaxies. We have restricted our analysis to the velocities ranging between 9,000 and 18,900 km/s, i.e. within a redshift range of $0.03<z<0.063$ corresponding to the spectrospic redshifts of galaxies belonging to the Shapley supercluster similarly to \cite{quintana20} (see also Fig. \ref{FigFoG_correction}). Furthermore following \cite{quintana20} assessment of the reliability of the photometric selection, we also used only the galaxies for which R2 magnitudes\footnote{Derived from the SuperCOSMOS scans of photographic Schmidt survey plates \url{http://www-wfau.roe.ac.uk/sss/index.html}, data cover BRI bands, with the R represented at two epochs R1 and R2.} are  brighter than 18. This amounts to a total of 5,553 galaxies.

In order to account for the complex filamentary structures in the 3D velocity distribution, we need to correct for effects such as Fingers of God \citep[FoG,][]{FoG-hist1972} due to clusters and groups of galaxies (see Fig. \ref{FigFoG_correction} upper panel) that represent stretched distribution of galaxies which could be mistaken for actual filaments of the cosmic web. We therefore implemented a method to statistically correct for these redshift distortions present in the selected sample of spectroscopic galaxies which we use in Sect. \ref{anal} to identify three-dimensional filamentary web around Shapley. 

The correction of FoG effect proceeds in two steps, first identifying the groups and clusters of galaxies and second compressing the elongated structures caused by FoG. The procedure we used to identify the groups and clusters follows the group finding algorithm described in \cite{duarte2014} and references therein. First, we identify galaxies within the same bound structures using a Friend-of-Friend (FoF) algorithm which depends on two linking lengths normalised to the mean nearest neighbour separation of field galaxies, along line-of-sight, los, ($b_{\mathrm{los}}=1.0$) and in the plane of the sky ($b_{\mathrm{plan}}=0.2$). 
Subsequently, we correct for the FoG effect. Namely, the radial distances of galaxies, identified as members of the structures by the FoF algorithm, are redistributed so that the redshift-distortion elongation along the los is compressed around the centre of the structure. We assume that the distribution of galaxies inside groups and clusters is symmetrical both perpendicularly and along the los \citep[see, e.g.,][]{2004ApJ...606..702T,2012A&A...540A.106T,2016ApJ...818..173H}. The compression factor is computed from the ratio between the rms of galaxies distance, w.r.t. the centre of the group or cluster, perpendicular and along the los. As a result, the correction to the FoG distortions slightly modifies the redshift range which is reduced to $0.03<z<0.058$.
\subsection{Gas tracers}
In addition to the galaxy distribution in the area of Shapley, we also used two tracers of the hot gas in SSC, namely the X-ray emission and the thermal Sunyaev-Zel'dovich (tSZ) signal \citep{Sunyaev1970}. 

For the X-ray emission, we used the publicly available data from the ROSAT survey \citep{snowden1997} which include count rates over the whole sky. The X-ray maps\footnote{\url{https://www.jb.man.ac.uk/research/cosmos/rosat/}} in HEALpix format \citep{2005ApJ...622..759G} are provided in six standard energy bands covering $0.1-2$ keV. For the present work, we used the three higher energy bands covering $0.56-2$ keV. 
For the tSZ signal, we made use of the new full-sky Compton parameter map ($y$-map) reconstructed by \cite{HidekitSZ} from the \textit{Planck} multi-frequency channel maps, between 100 and 857~GHz, within the \textit{Planck} data release 4 \citep{2020A&A...643A..42P}. 

In both full-sky maps, we extracted a patch of $15\times 15$ square degrees centred on Shapley and corresponding to the whole field. We display, in Fig. \ref{FigFull_map},  contours illustrating inner and outer regions of the SSC in the tSZ and X-ray signals. The outer and inner contours of the tSZ signal correspond to Compton $y$ parameters values of $4.4\times 10^{-6}$ and $1.1\times 10^{-5}$, respectively. The outer and inner contours in the X-ray signal correspond to 2000 counts$\,$s$^{-1}\,$arcmin$^{-2}$ and $10^{4}$ counts$\,$s$^{-1}\,$arcmin$^{-2}$, respectively. 
%
%
\section{Analysis} \label{anal}

   \begin{figure*}
   \centering
   \includegraphics[width=0.45\textwidth]{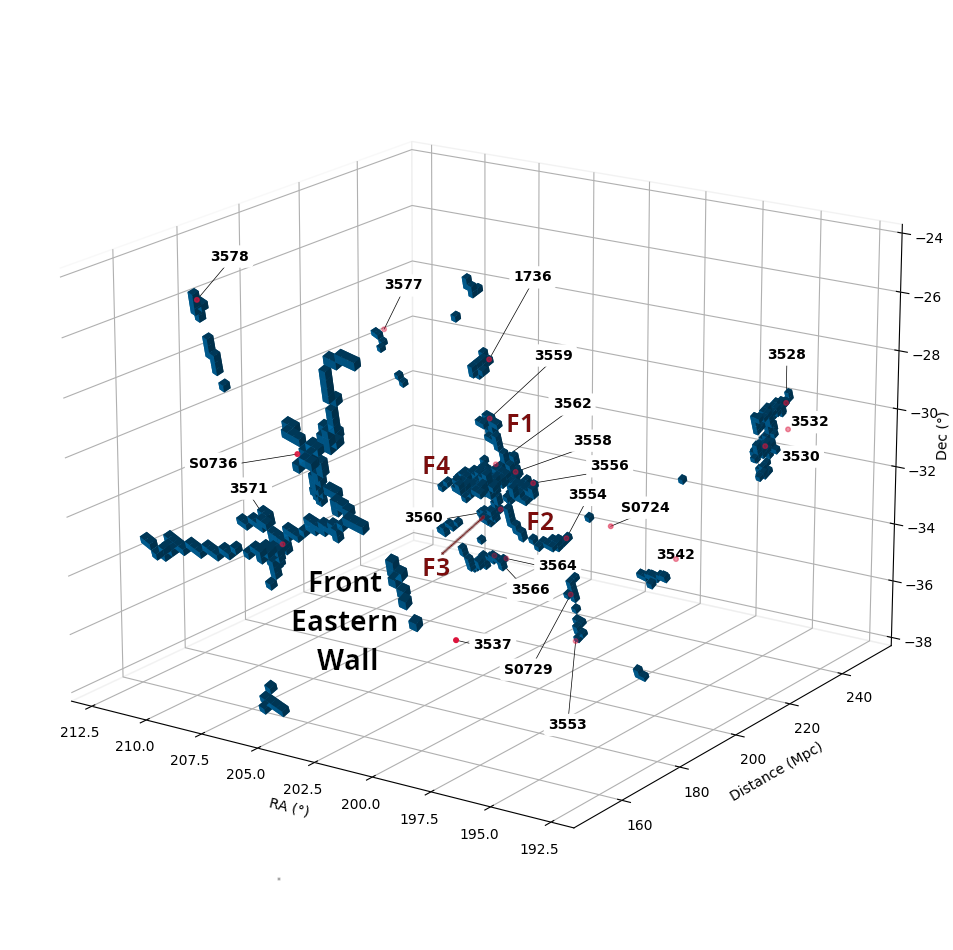}
   \includegraphics[width=0.45\textwidth]{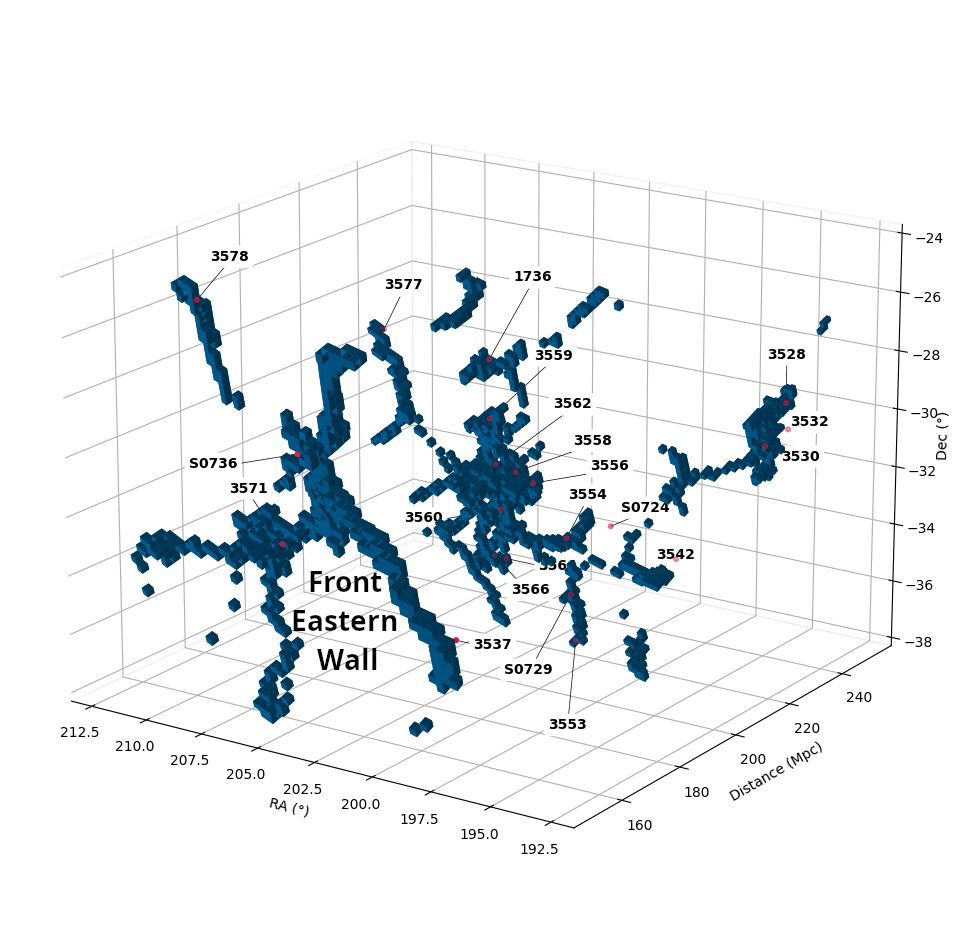}
      \caption{3D visualisation of the T-ReX filaments around SSC, with a probability above 0.5 (left), defined as our baseline for reliable filaments, and a lower probability of 0.2 (right). The green dots correspond to the clusters from \citet{Proust2006A&Acatalogue}.
              }
         \label{FigFilament_3D}
   \end{figure*}
%
   \begin{figure*}
   \centering
   \includegraphics[width=0.45\textwidth]{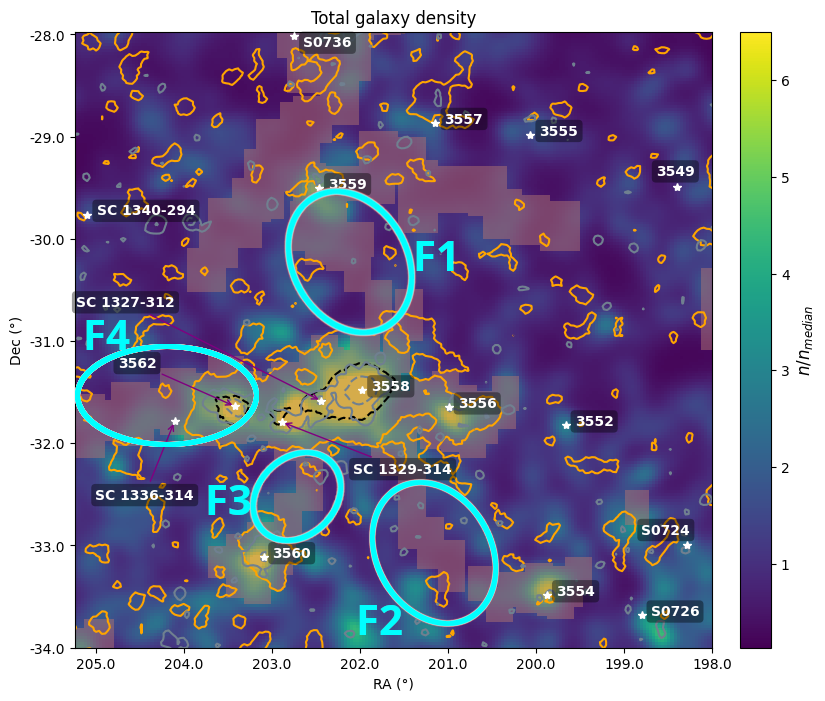}
   \includegraphics[width=0.45\textwidth]{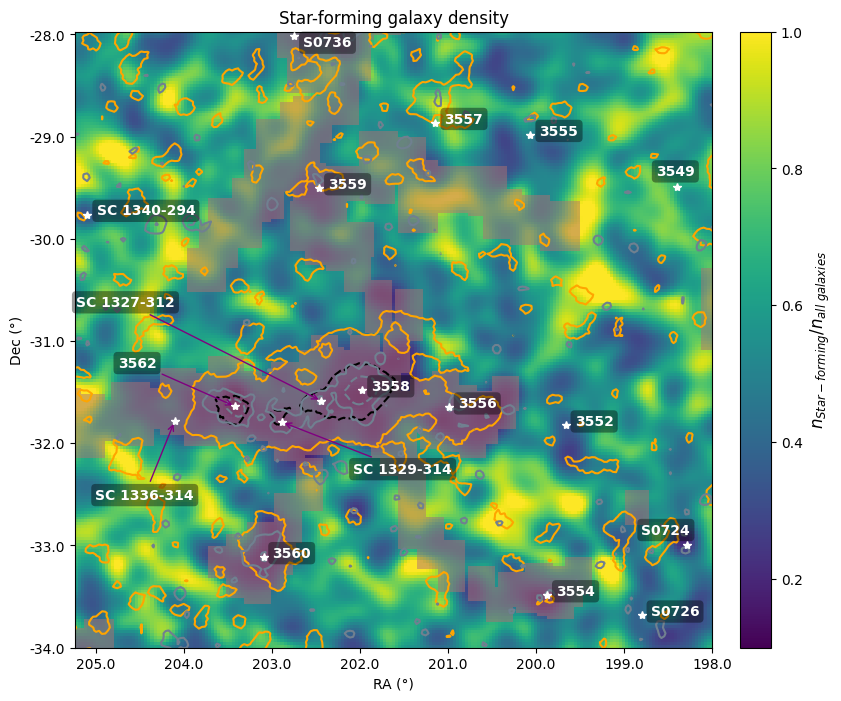}
   \includegraphics[width=0.45\textwidth]{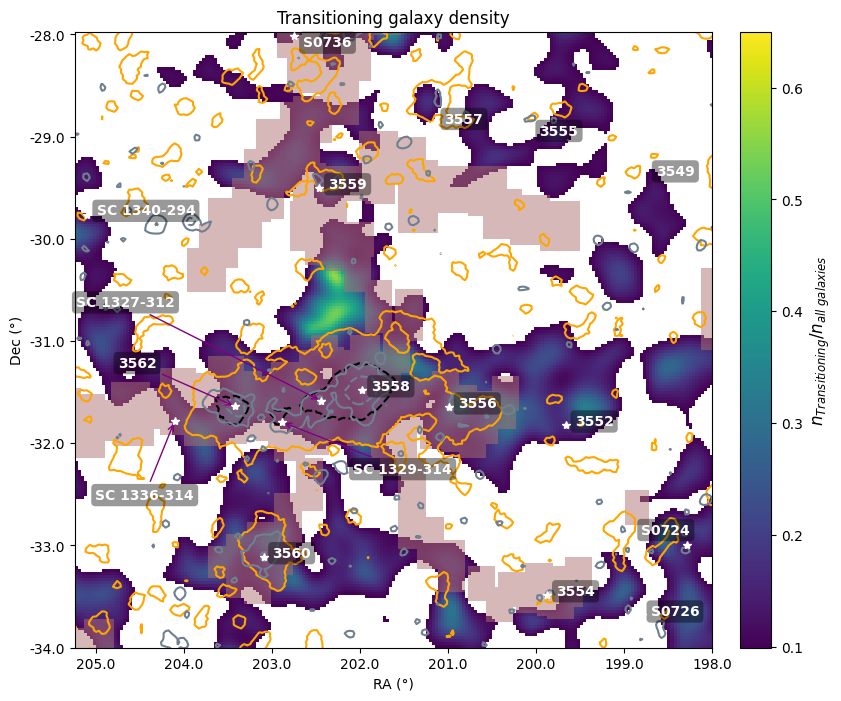}
   \includegraphics[width=0.45\textwidth]{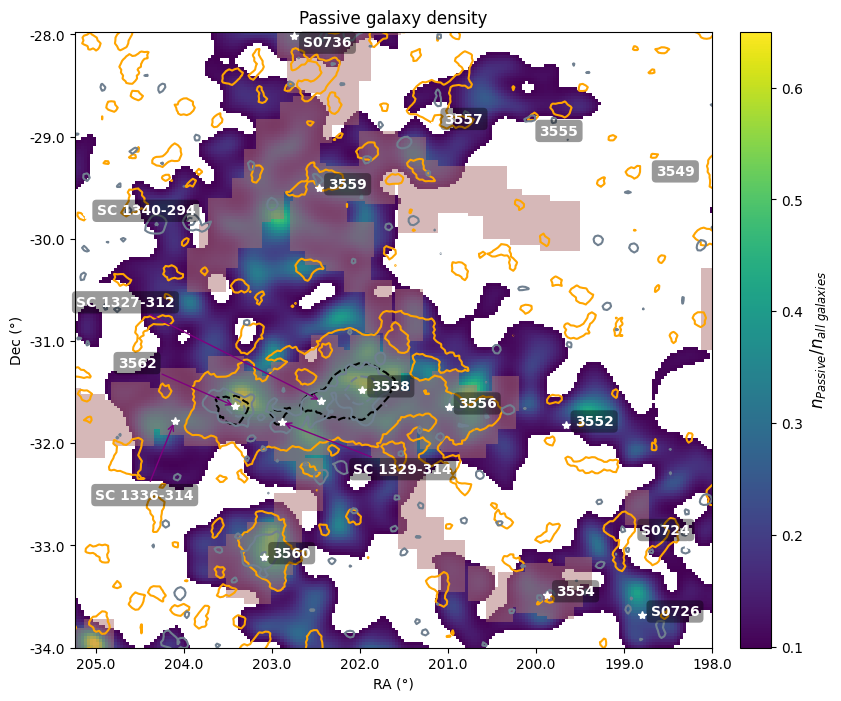}
   \caption{\textit{Upper left}: Same as Fig. \ref{FigFull_map}, with T-ReX filaments shown in gray areas. Main detected filaments are indicated with cyan ellipses. \textit{Upper right}: Zoom-in region around the SSC system with normalised number density of star-forming galaxies. \textit{lower right and left}: Same as upper right for transitioning and passive galaxies, respectively. The white areas correspond to normalised number density values such that $n_\mathrm{population}/n_\mathrm{all/ galaxies} < 0.1$.
   White stars correspond to known clusters, projected 3D filaments are indicated by gray areas, while orange and turquoise contours correspond to the tSZ and X-ray signals, respectively. 
   }
              \label{FigPopulations}%
    \end{figure*}
\begin{figure*}
\centering
\includegraphics[width=0.45\textwidth]{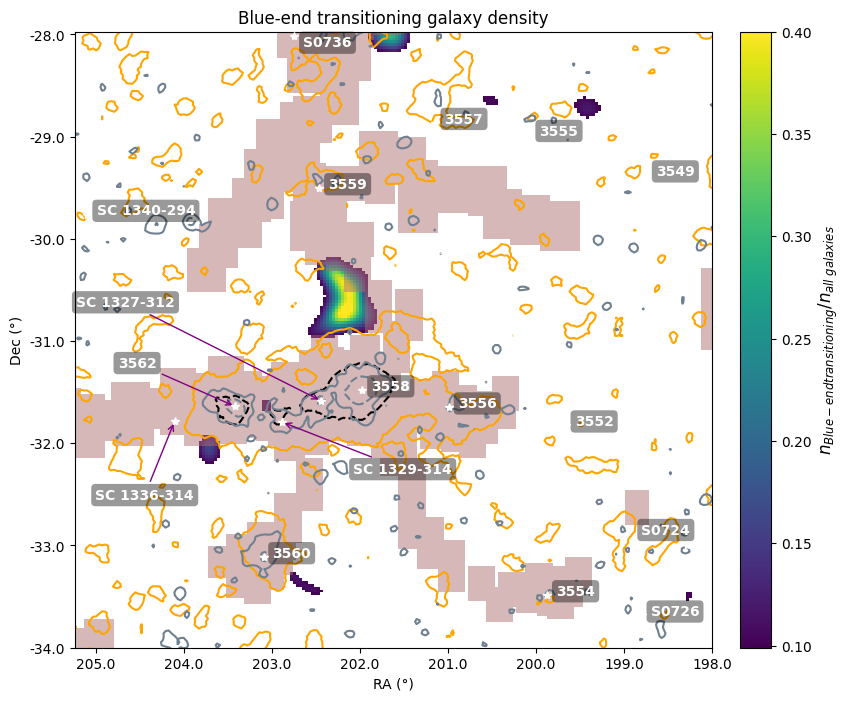}
\includegraphics[width=0.45\textwidth]{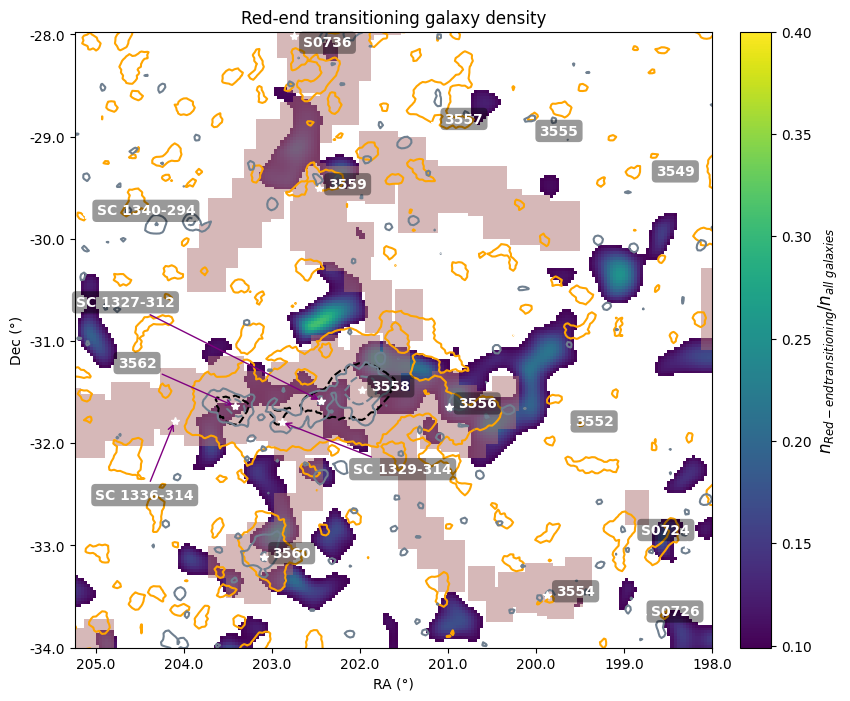}
\caption{Same as lower left panel in Fig. \ref{FigPopulations}, with the blue-end transitioning galaxies (lower d2ms hence more star-forming) on the left and the red-end transitionning galaxies (higher d2ms hence more passive) on the right. The colour bar is adjusted to better highlight the differences. As in previous figures, white stars are known clusters and the gray areas correspond to detected filaments.
    }
     \label{FigTransitioning}%
\end{figure*}

\subsection{Filament detection} \label{fil-dect}
Elongated structures connecting galaxy clusters were seen in the observations of Shapley supercluster \citep[e.g.,][and references therein]{quintana20,higuchi2020}. In this study, we use for the first time an actual filament detection technique to identify the filaments in the SSC area. We perform the filament detection both in 3D, using the spectroscopic galaxy sample corrected for the FoG effects, and in 2D, using the galaxies from the WISExSCOS photometric redshift catalogue. The detection in 3D provides the actual filaments connecting the clusters of the SSC area. The detection in 2D shows what is accessible when limited by the precision of the photometric redshifts.

Several web-finder algorithms have been developed to identify the environments of the cosmic web  \citep[voids, walls, nodes, and filaments, see, e.g., review by][]{libeskind2018}. Here, we choose to use the Tree-based Ridge Extractor (T-REx) algorithm \citep{Bonnaire2020,2021arXiv210609035B} in order to detect the filamentary network in the SSC area, \citep[see][for a comparison with other filament finders applied on simulations]{Bonnaire2020}. T-REx models the filamentary pattern as a graph structure and outputs a smooth version of the minimum spanning tree using a Gaussian mixture model to describe the spatial distribution of the galaxies. Hence, one of the advantages of T-REx is that it directly estimates the filament spines from the galaxy distribution without resorting to the construction of smooth density field. Moreover by using graphs to trace interconnected galaxies, T-REx does not assume any shape or width for the filaments that can rather be learnt during the optimisation.

In order to extract a reliable 3D filamentary network from the galaxy distributions, it is essential to use the radial distances of the galaxies. We thus used the distances from the spectroscopic galaxies sample corrected for FoG effects (see Sect. \ref{specGal}). The output of T-REx is filaments represented by graph ridges and indicating one particular path linking graph nodes  together, hence only one possible spine, without providing any idea of uncertainty of the linking graph. In order to estimate the reliability of the detected filaments, we follow the approach of \cite{Bonnaire2020} who introduced a robust representation that takes into account the possible variations in the input node distribution by defining and analysing random subsamples of the input distribution. In a similar fashion as in bootstrap approaches, we perform 100 iterations of the filament extraction. In practice, the input data is divided into 100 randomly selected subsamples, each containing $75\%$ of the galaxies from the original sample as shown in \cite{Bonnaire2020}. In that way, each subsample traces one specific linking path, that is, a specific set of filaments. Subsequently, we  construct a 3D probability grid based on the number of times a given cell is crossed by a realisation of a filament. This probability is represented by a fractional value between 0-1, with 0 for no filament out of 100 samples detected in a given cell, and increasing probability of detecting filaments from individual realisations up to a value of one for a cell where every realisation has recovered a filament. The significance of the detected filaments can be defined by setting a threshold value between 0 and 1. A conservative limit, close to 1, may result in filaments that appear disconnected, while a lower limit, close to 0, may include spurious filaments (see Fig. \ref{FigFilament_3D} for a comparison between two threshold probabilities 0.2 and 0.5). In the following we trade reliability and completeness, out of all the realisations of the filamentary pattern detected by T-REx we focus only on the reliable ones, namely we select the cells with a probability above 0.5 (see Fig. \ref{FigFilament_3D} left panel). A projection of the 3D cells containing these reliable filaments is shown as a gray area in Figs. \ref{FigPopulations}, and \ref{FigFilament_comparison}. 

Despite the large uncertainties on the photometric redshifts, we also applied the T-ReX algorithm to the subset of passive galaxies, which trace the filaments \citep{bonjean2020}, from the photometric galaxies sample described in Sect. \ref{photGal}. This provided us with a catalogue of filaments projected on the sky.  Within the SSC region, the 2D and 3D filamentary patterns are overlaid (Fig. \ref{FigFilament_comparison}) showing the overall agreement between the 3D and 2D structures of SSC.

\subsubsection{The SSC structure}\label{sec:struc}
\begin{figure}
   \centering
   \includegraphics[width=\hsize]{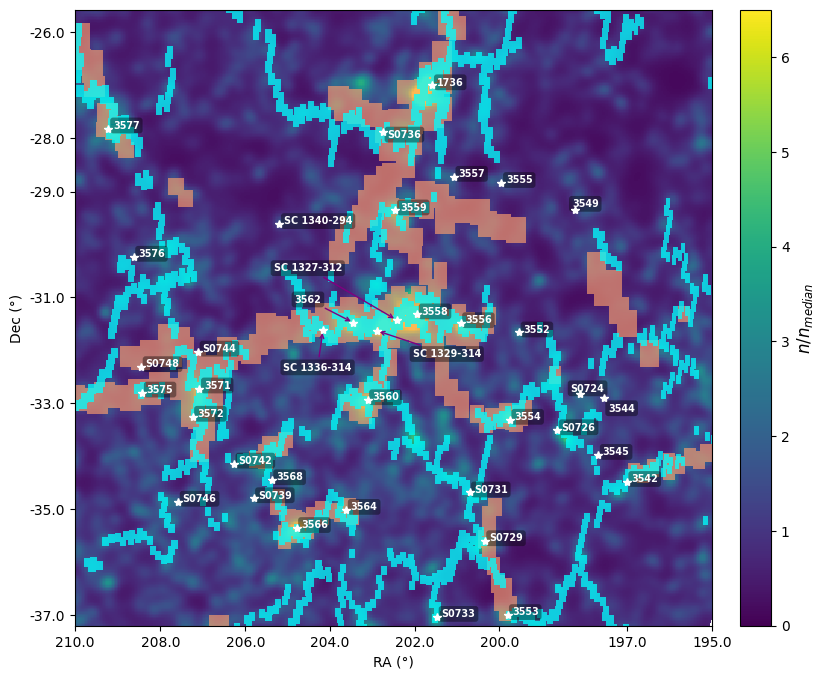}
      \caption{Same as Fig. \ref{FigFull_map}, with the filaments overplotted on the galaxy density map. The filaments were obtained in 2D from the passive population in the WISExCOSMOS added value catalogue (cyan) and in 3D from the spectroscopic catalogue (brown areas). The 2D and 3D filaments have a probability above 0.5, as given by the T-ReX algorithm. The black stars correspond to clusters from \citet{Proust2006A&Acatalogue}.
              }
         \label{FigFilament_comparison}
   \end{figure}

The overall filamentary structure of the SSC as traced by the galaxy overdensity has already been presented and discussed in details by \cite{2006Msngr.124...30P} and \cite{quintana20} and references therein. In the following, we present the results of a blind detection of the filamentary pattern by the T-REx algorithm applied on FoG corrected galaxy positions. We show how our results, when considering reliable filaments only (probability above 0.5), compare with those of previous studies and how they provide a renewed view of the SSC global structure. 
\begin{itemize}
    \item Applying the web-finder, T-REx, on the spectroscopic galaxy sample, we clearly detect an extended and reliable (probability above 0.5) filamentary pattern (Figs. \ref{FigFilament_3D} and \ref{FigFilament_comparison}) relating the central region of the SSC as defined in \cite{quintana20} and references therein, namely the groups and clusters of galaxies A3552, A3554, A3556, A3558, A3559, A3560, AS0724, AS0726, SC1327-312, SC1329-313. With T-REx, we identify very large scale filaments. We identify the filament in the North connecting the core region of SSC to the cluster A3559 (via Filament 1, labelled F1 in Figs. \ref{FigFilament_3D} and \ref{FigPopulations}). This elongated structure extends even further. We find a filament extending from the core of SSC beyond A3559 in the overall direction of A1736 to which it is connected with lower reliability filament identified with T-REx. We clearly note the T-REx high reliability filament in the South of SSC relating the core of Shapley and the clusters of galaxies A3560 (via Filament 3, noted F3), A3562 (via Filament 4, F4), and 3554 (via Filament 2, F2). An elongated structure extends towards A3566 to which it is connected with lower reliability filament.
 
    \item Foreground galaxies form “Front Eastern Wall” identified in
    \cite{quintana20} which extends to the east and in front of SSC, composed of a bridge of galaxies, groups and clusters. With T-REx, we confirm the detection of such a very large complex with filaments connecting A3578 in the North to A3537 in the South and linking the groups and clusters A3571, A3575, S0744, and S0736 (Fig. \ref{FigFilament_3D}). Considering slightly lower reliability filaments detected with T-REx, the complex network includes the groups and clusters SC1336-314  A3570, and S0748. 
    
    \item We also detect with T-REx and hence confirm the large scale structure in the North West of the SSC core made of the clusters of galaxies A3528, A3530 and the filaments connecting them, in addition to a lower reliability T-REx filament connected to A3532. The web-finder algorithm identifies several realisations of filaments bridging between this structure and the SSC. However, only a portion of the filaments pass our reliability threshold (probability above 0.5). 

\end{itemize}

\subsection{Galaxy populations in the SSC environment} \label{sec:SSCenv}

We examine the spatial distribution of the galaxy types in the SSC region. We compare it against the spatial distribution of the hot ionised gas, traced by the tSZ and X-ray signals. We also identify a zoom-in region of $7 \times 6$ deg$^{2}$ centred in the Shapley core in order to avoid the foreground structures from the SSC larger field (see Fig. \ref{FigFilament_3D}). In Fig. \ref{FigPopulations}, we display the quantitative fractions of galaxies from each population and overlay the contours of the tSZ and X-ray signals together with the areas defined by the projected 3D filaments. 

As expected from the quenched nature of galaxies in clusters \cite[see][for a seminal study]{dressler1980}, we clearly observe that the passive galaxies occupy the densest environments of SSC corresponding to clusters and groups where the tSZ signal is the largest. This is particularly well seen in Fig. \ref{FigPopulations} (also in Fig. \ref{FigFull_map}) for example at the positions of isolated clusters of galaxies in the field such as A3554, A3556, A3558, A3560, A3562, A3566 or A1736 where we observe consequently at the same positions that theses regions are devoid of star-forming and transitioning galaxies. All over these regions, the fraction of the smoothed number densities of passive galaxies is above 0.4 as compared to values below 0.2 in the background region. In the very same regions, the fraction of star-forming and transitioning galaxies remains below approximately 0.2. \cite[see][for a seminal study]{dressler1980}. \\
Not all the clusters in the SSC seem to have reached a fully quenched galaxy stage. In the very core of the SSC where the tSZ signal is the highest, we additionally observe a large density of star-forming and transitioning galaxies, in addition to the passive population as seen in the right panel of Fig. \ref{Fig:d2ms_histogram}.. This observation is the signature of a significant star-formation activity that could be triggered by gas supplied in the filaments or by cluster interactions or by nearby groups or field galaxies. For example, extended tSZ emission in the region North SC1329-314 and SC1327-312 is filled with transition galaxies with contribution from  star-forming galaxies and just very few passive galaxies (see also Fig. \ref{FigTransitioning}). This observation is coherent with the observations on the clusters in the very core region of SSC. Indeed, \cite{merluzzi2015} states that A3558 is considered as a late merger-stage cluster while \cite{quintana20} proposes that the groups SC1329-314 and SC1327-312 could be associated to the in-fall of matter along the line of site.
\\
Outside the core region of SSC, we also find clusters dominated by a population of transitioning galaxies, as in the case of A3559, or clusters with the core dominated by passive galaxy population and surrounding shells of transitioning and star-forming galaxies, such as AS 0726 or A1736. In this latter case, the overdensity of transitioning galaxies at the north side of the cluster is overlaping with the peaks of the X-ray and tSZ signals.

Focusing on the large-scale spatial distribution of hot gas as traced by the tSZ contours in Fig. \ref{FigPopulations}, we clearly observe an extended tSZ signal north of A3558 associated with the location of an overdensity of transitioning galaxies. In this region where a filament was detected with T-REx, the fraction of the smoothed number density of passive galaxies exceeds a value of 0.5 (see lower left panel in Fig. \ref{FigPopulations}), while there is no significant excess of  star-forming or of passive galaxies. This region, dominated by transitioning galaxies, is remarkably well delineated by the outer tSZ contour in the South and the projected T-REx filament in the West. This would imply that galaxies are impacted by the filament and the supercluster, and being quenched by the higher density and the gas temperature in these environments. Moreover, we found that the transitioning galaxies with lower d2ms values (i.e., the most star-forming galaxies among the transitioning population) are pressed towards the filament, while the galaxies with higher d2ms (i.e., the most passive galaxies among the transitioning population) are closer to the tSZ contour (see Fig. \ref{FigTransitioning}). Indeed, in this latter region we also find a significant fraction of passive galaxies. This spatial distribution suggests that either the supercluster environment is more efficient at quenching the galaxy star formation, or this quenching started earlier closer to the SSC than to the filament. Similarly, the extended regions of tSZ signal west of A3556 and north of A3562 seem associated with overdensities of transitioning galaxies, situated along the filaments detected with T-REx, with the passive galaxies being at the locations of the cluster centers. 

%
   \begin{figure}
   \centering
   \includegraphics[width=0.45\textwidth]{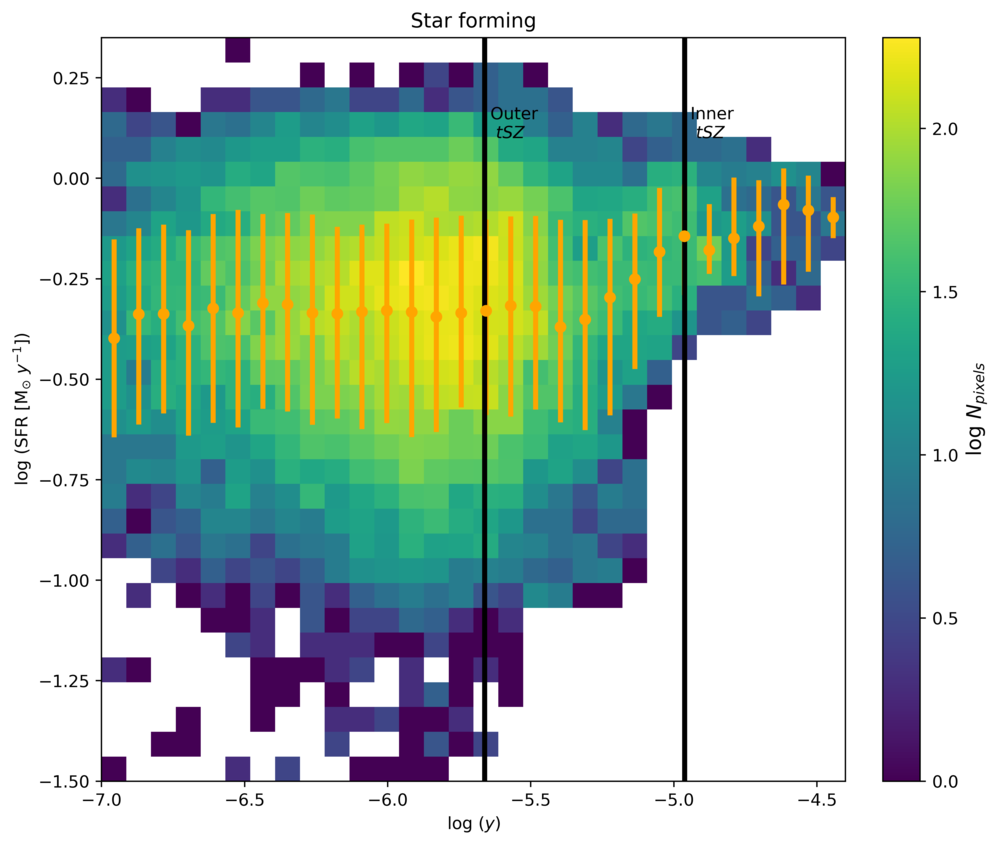}
   \includegraphics[width=0.45\textwidth]{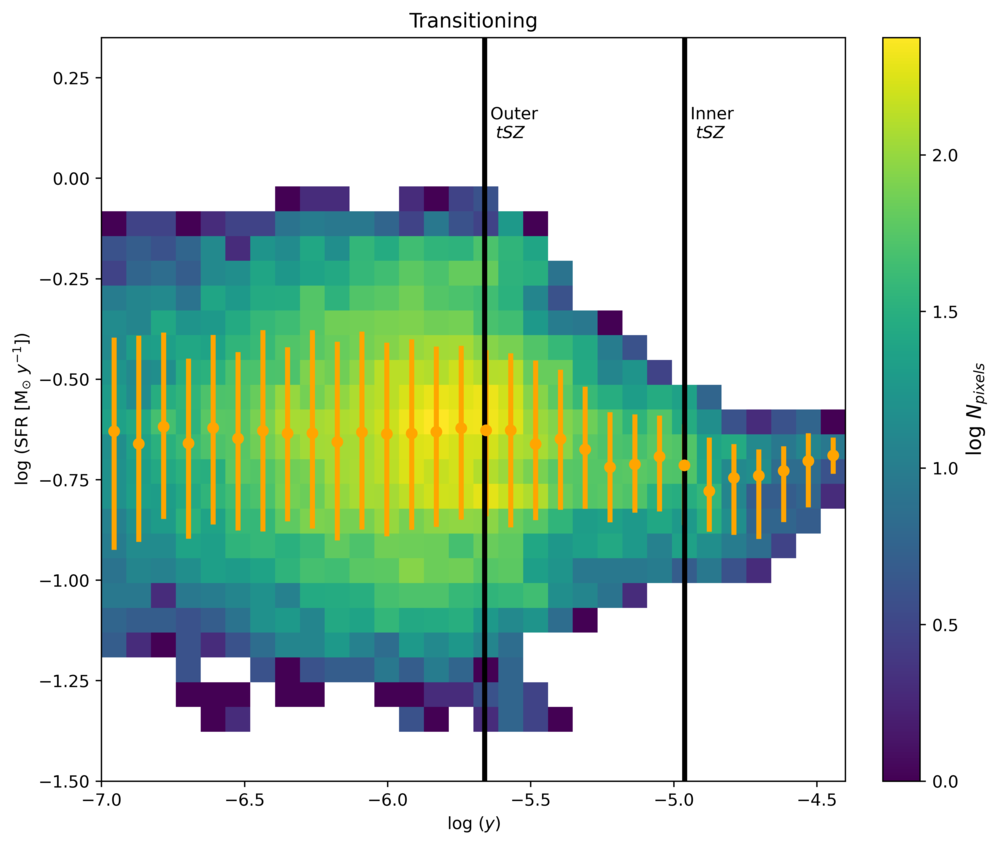}
   \includegraphics[width=0.45\textwidth]{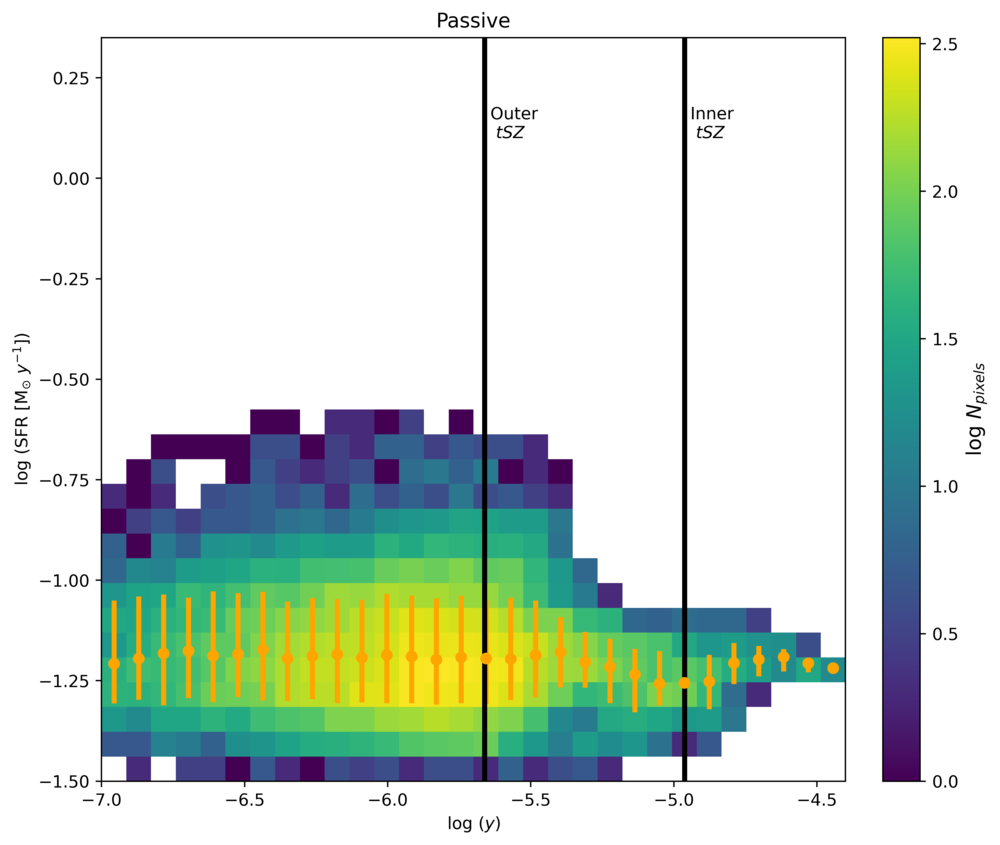}
   \caption{SFR and Compton $y$ relation for star-forming (upper panel), transitioning (middle panel) and passive galaxies (lower panel) in the zoom-in region of SSC. The colour scale indicates the number of pixels for each combination of SFR and $y$, while the dots and bars are the median and 68\% distribution, respectively. Vertical black lines correspond to the same tSZ outer (orange solid, $y_\mathrm{outer} = 2.2\times10^{-6}$) and inner (black dashed, $y_\mathrm{inner} = 1.1\times10^{-5}$) contours in Fig. \ref{FigPopulations}. }
              \label{FigSZ_SFR}%
    \end{figure}
%
\subsection{SFR-tSZ relation}\label{SFR-y}
The results, presented in Sect. \ref{sec:SSCenv}, illustrate the interplay between galaxy evolution and the environment (clusters, filaments) in which they are located. The latter not only contains the galaxies but is also filled with gas, presumably hot ($T_\mathrm{e}>10^6$~K), traced by tSZ or X-ray signals. In the rest of the analysis, we investigate the environmental effects within the filamentary pattern of SSC on star formation in galaxies, traced by the SFR. On the one hand, we make use of the SFR maps constructed for passive, transitioning and star-forming galaxies (described in Sect. \ref{photGalmap}). On the other hand, we use the tSZ map providing us via the Compton $y$ parameter values a measure of the pressure of the gas in the SSC field and hence defining in a single quantity, which combines both density and temperature, the environment in which galaxies evolve. Focusing on the zoom-in region of $7\times6 $ square degrees shown in Fig. \ref{FigPopulations}, we compute the SFR in the pixel maps for each of the galaxy populations and display this quantity against the Compton $y$ value of the tSZ map (see Fig. \ref{FigSZ_SFR}). To avoid background noise in the tSZ map we consider only pixels above a threshold value of $y = 10^{-7}$. The vertical lines, from left to right at $y_\mathrm{outer} \approx 2.2\times10^{-6}$ ($\log y_\mathrm{outer} \approx -5.66$) and $y_\mathrm{inner} \approx 1.1\times10^{-5}$  ($\log y_\mathrm{inner} \approx -4.96$), indicate the Compton parameter values associated with tSZ signal contours delimiting the outer region, i.e. where tSZ is dominated by background, and the inner core region of the SSC, i.e. dominated by the tSZ signal from hot gas. In Fig. \ref{FigSZ_SFR}, the colour indicates the number pixels with the corresponding SFR and $y$. The orange dots and associated error-bars indicate the median SFR at a given $y$ and the 68\% dispersion around the median, respectively.  \\
As expected, we find that the star-forming galaxies have a larger median SFR ($\approx 0.46$ M$_{\odot}$ $y^{-1}$) than the transitioning  ($\approx 0.23$ M$_{\odot}$ $y^{-1}$) which are more star-forming that the passive galaxies ($\approx 0.06$ M$_{\odot}$ $y^{-1}$). 
Turning to the slopes of the SFR-$y$ relations, deviations from a zero slope indicate that the SFR is impacted by the tSZ signal from hot gas. In the region defined by the outer contours ($y<2.2\times10^{-6}$) independently of the galaxy type, we observe no departure from zero-slope suggesting that there is no impact of the tSZ signal on the SFR in galaxies apart from the value of the median SFR. In the SSC region defined by the tSZ contours ($y>2.2\times10^{-6}$), the transitioning galaxies show mostly a zero-slope relation whereas star-forming galaxies experience an increase of their SFR above a Compton parameter value $y_q >3.4\times10^{-6}$ ($\log y_{q} > -5.46$, where the SFR-$y$ slope $>0$). This could be explained by enhanced gas compression triggering an increase of star formation activity, as suggested in previous studies \citep[e.g.,][]{rawle2014,castignani2020}. This increase coincides with a small decrease of the SFR as a function of $y$ in the passive galaxies. We show in Table \ref{table:Spearman} the results of the Spearman test run over the pixels of the maps (sample size $\approx 10,000$) for the correlations between Compton $y$ parameter and SFR and galaxy number density. All $p$ values were found smaller than $10^{-6}$.

   \begin{figure*}
   \centering
   \includegraphics[width=0.45\textwidth]{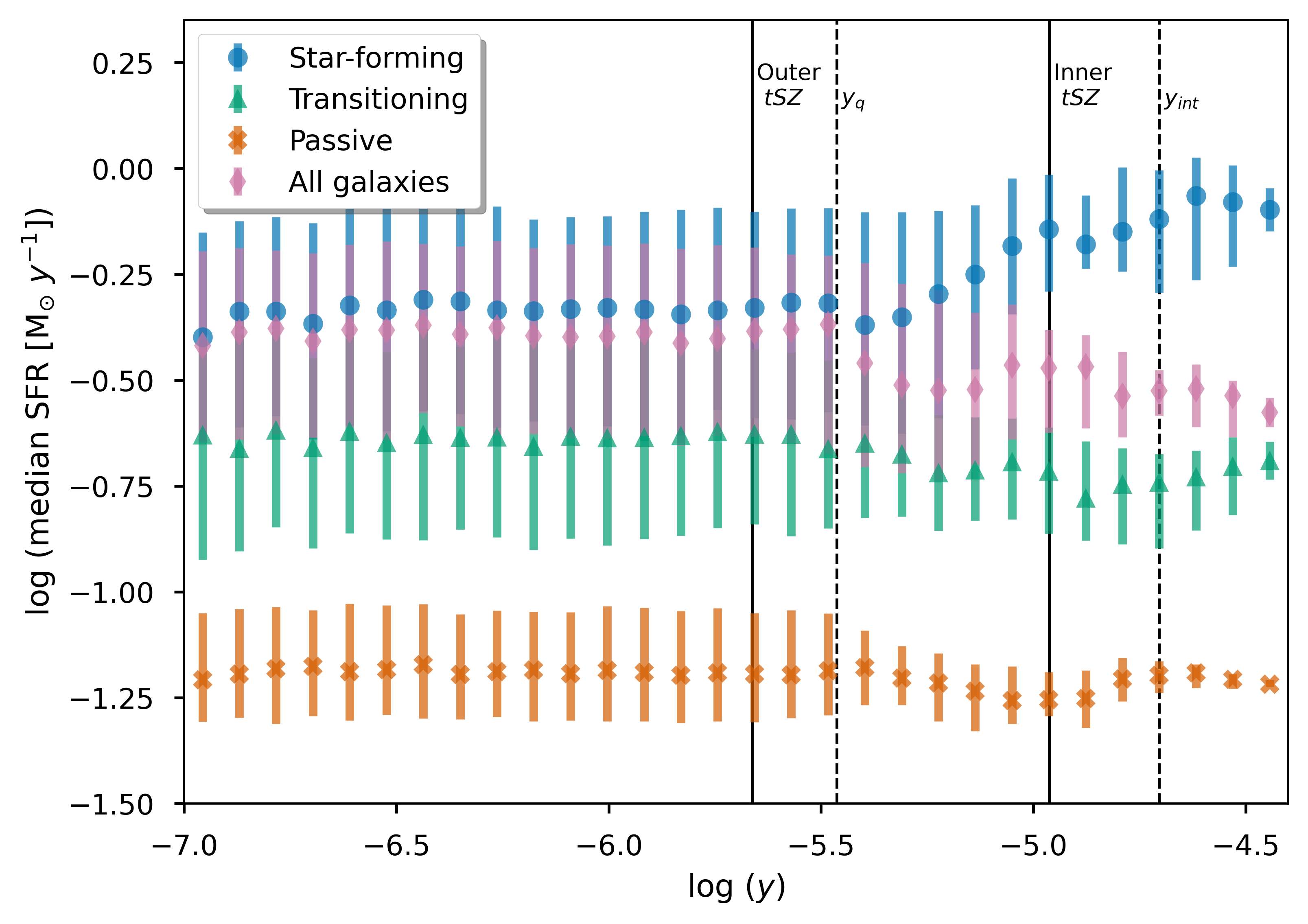}
   \includegraphics[width=0.45\textwidth]{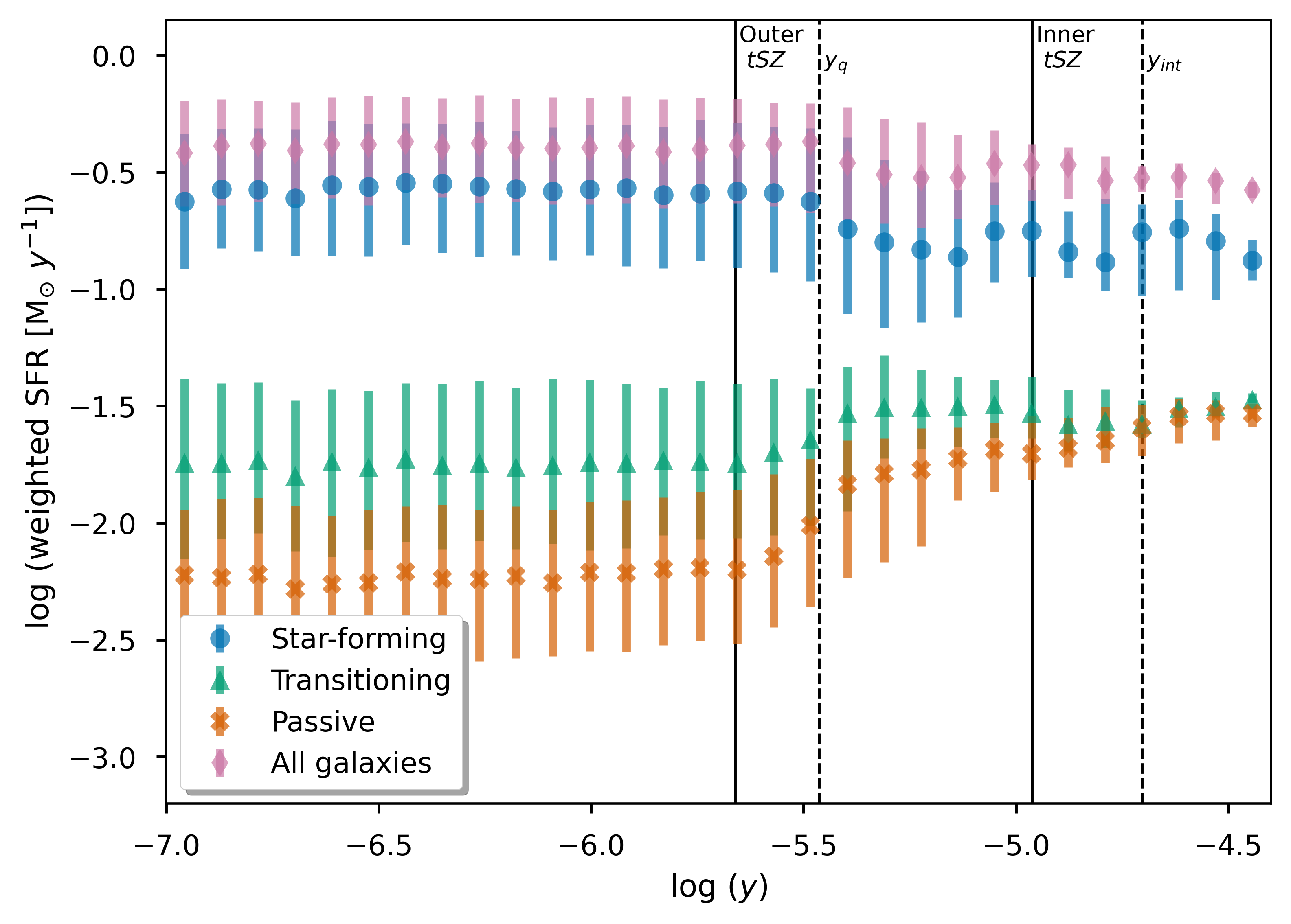}
   \caption{\textit{Left panel:} Median SFR values for all three galaxy populations (star forming, transitioning and passive) as a function of the tSZ $y$ parameter. \textit{Right panel:} Distribution of the median SFR weighted by the fraction of galaxy-density of a given population w.r.t. total as a function of the tSZ $y$ parameter. In both panels,
   error bars indicate the 68\% distribution around the median. Vertical solid lines correspond to the tSZ outer ($y_\mathrm{outer}=2.2\times10^{-6}$) and inner ($y_\mathrm{inner}=1.1\times10^{-5}$) contours. Vertical dashed lines represent Compton parameter values at which the slopes of the SFR-tSZ relation significantly change (see text).}
              \label{FigSZ_SFR_medians}%
    \end{figure*}

\bigskip
The number density of galaxies of different types are different. Hence in order to take this into account in our interpretation of the star-formation activity vs tSZ in Fig. \ref{FigSZ_SFR_medians} (left panel), we compute the SFR weighted by the fraction of galaxy of a given type relative to the total number of galaxies (see Fig. \ref{FigPopulations} for the associated maps). We display the results in Fig. \ref{FigSZ_SFR_medians} (right panel) where we also show the full galaxy sample in black. We observe, as expected, that it is the sum of the contributions from the three galaxy populations relative to their number density. We can focus on the different regimes of tSZ signal defined by the contours of the Compton $y$ map.

\begin{table}
\caption{Spearman correlation coefficients relating Compton $y$ to SFR, upper part of the Table (Fig. \ref{FigSZ_SFR}); and Compton $y$ to galaxy number density lower part of the Table (Fig \ref{Fig_nden_SFR_medians}), for the full range of $y$-values and for $\hat{y}_{q} < \log y < \hat{y}_{\mathrm{inner}}$ (with $\hat{y}_{q}=\log y_{q}=-5.46$ and $ \hat{y}_{\mathrm{inner}}=\log y_{\mathrm{inner}}=-4.96$).  All $p$ values were found smaller than $10^{-6}$.}             
\label{table:Spearman}      
\centering                          
\begin{tabular}{c c c c}        
\hline\hline                 
 \multicolumn{4}{c}{\textbf{$\log y$ versus SFR}}  \\    
    &  &   &    \\
    $y$ range & Star-forming & Transitioning & Passive \\    
\hline                        
   $\log y > -7$ & 0.08 & -0.06 & -0.07  \\      
   $\hat{y}_{q} < \log y < \hat{y}_{\mathrm{inner}}$ & 0.25 & -0.12 & -0.3 \\
    &  &  &   \\
\hline                        
\hline                        
 \multicolumn{4}{c}{\textbf{$\log y$ versus number density}}   \\    
 &   &  & \\
 $y$ range  & Star-forming & Transitioning & Passive \\    
\hline                        
$\log y > -7$ & -0.37  & 0.23 & 0.36  \\
   $\hat{y}_{q} < \log y < \hat{y}_{\mathrm{inner}}$ & -0.45 & 0.21 & 0.48 \\ 
\hline                                   
\end{tabular}
\end{table}

\begin{figure*}
\centering
\includegraphics[width=0.45\textwidth]{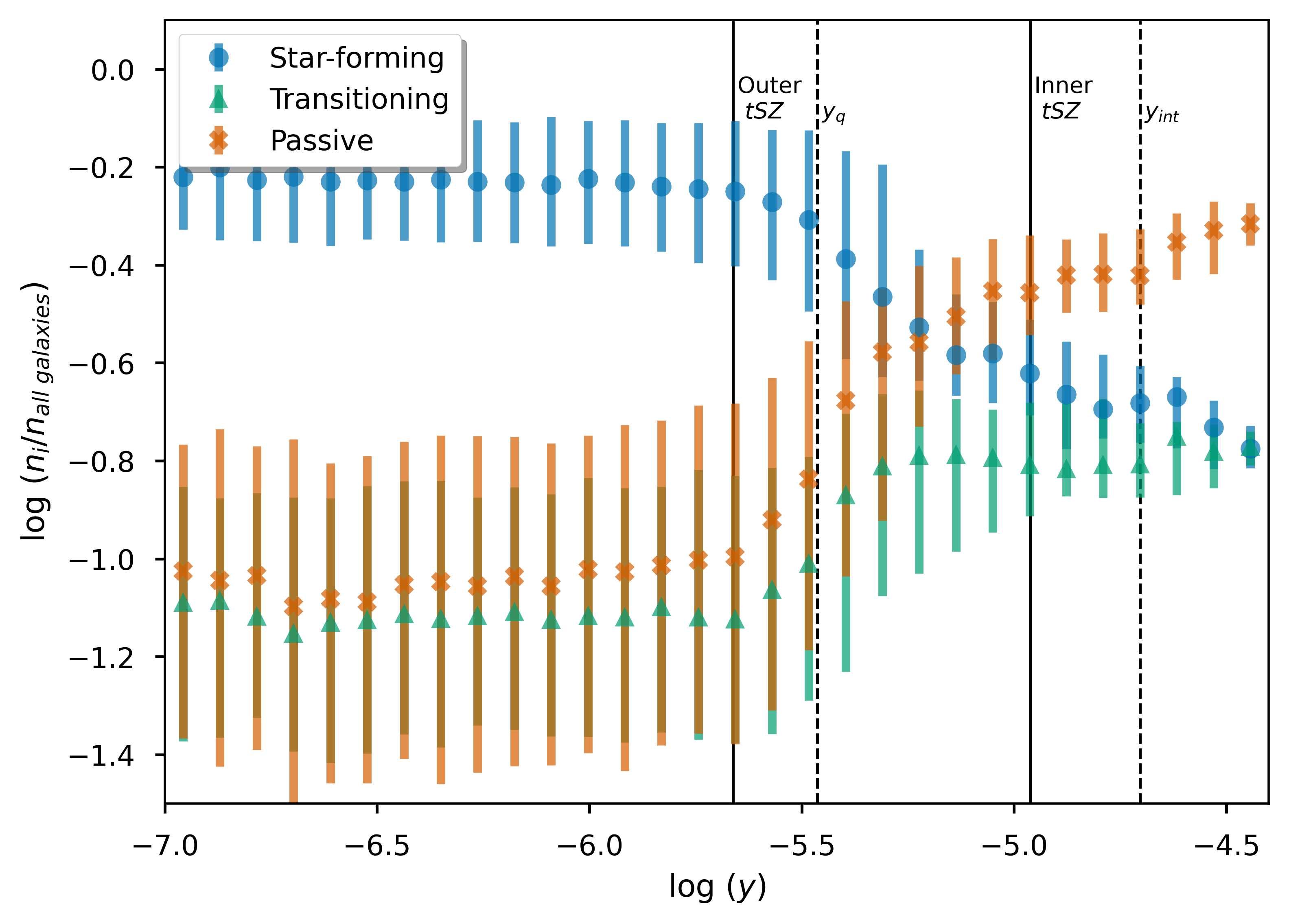}
\includegraphics[width=0.45\textwidth]{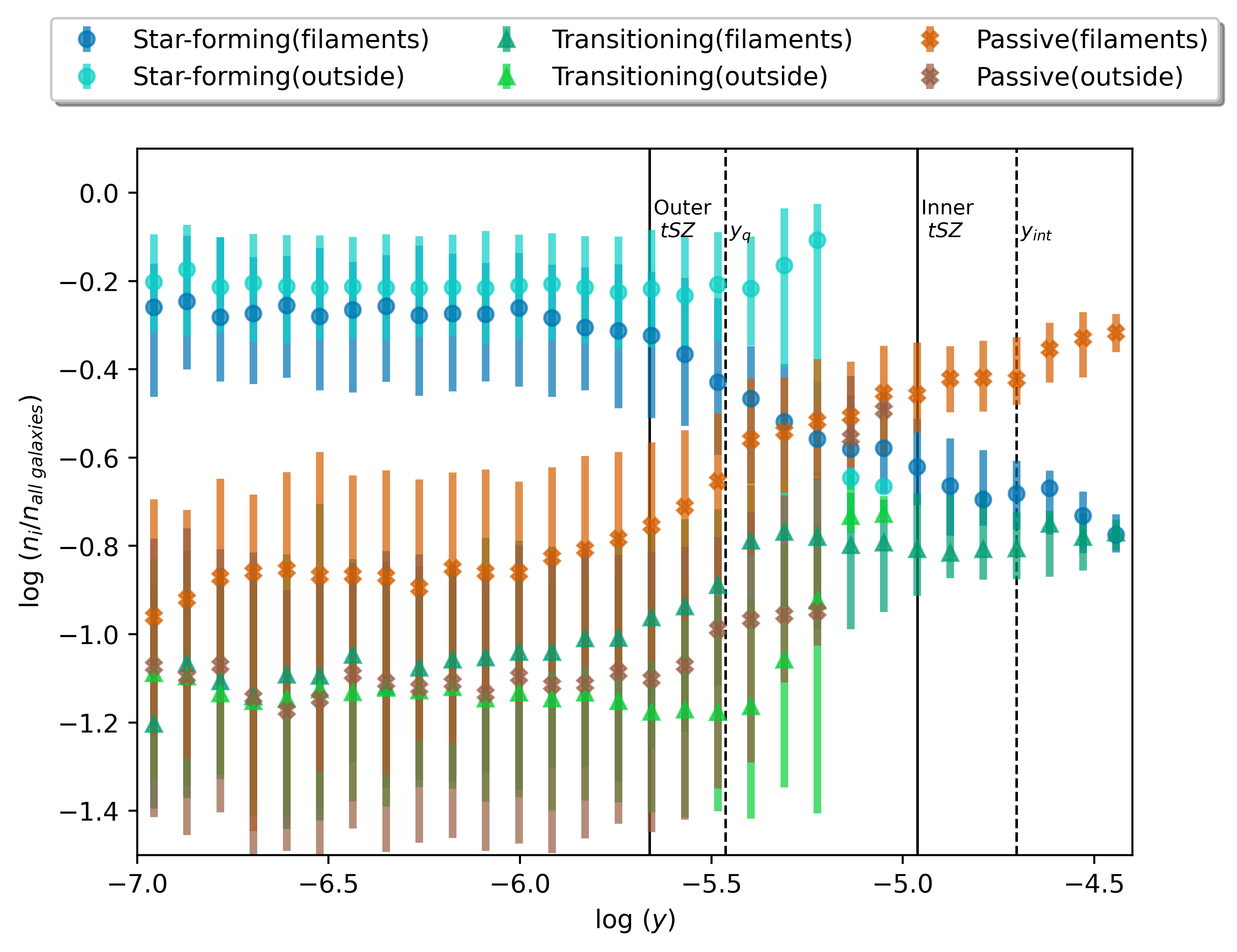}
\caption{\textit{Left panel:} Fraction of galaxy number density per population as a function of tSZ. Indicates the values by which the left panel in Fig. \ref{FigSZ_SFR_medians} was weighted to obtain the right panel in the same Figure. \textit{Right panel:} Same as left, but for galaxies inside and outside filaments, corresponding to Fig. \ref{FigSZ_SFR_filaments} }
\label{Fig_nden_SFR_medians}%
\end{figure*}

In the tSZ outer-contour region ($y<2.2\times10^{-6}$), background regime, we observe no impact of the tSZ effect on the SFR whatever the galaxy type. In the intermediate zone between the outer tSZ contours and the innermost contours ($2.2\times10^{-6}<y<1.1\times10^{-5}$), we notice a transition in the SFR-$y$ relation for all populations. At the Compton parameter value $y_{q} \approx 3.4 \times 10^{-6}$, the weighted SFRs of the total population and star-forming galaxies drop, while the weighted SFRs of transitioning and passive galaxies increase, already starting from the outer tSZ contour ($y>2.2\times10^{-6}$). The transition observed for the star-forming population suggests that the galaxies become quenched, at $y_{q}$, which translates in an increase of the number density of transitioning and passive galaxies at the expense of the star-forming galaxies. This increase is more significant given the slight decrease in SFR for transitioning and passive galaxies beyond $y_q$ (Fig. \ref{FigSZ_SFR_medians}, left panel). 
By contrast for the star-forming galaxies and as seen in Fig. \ref{FigSZ_SFR_medians} (left panel), the $y_{q}$ value coincides with an increase in the SFR. This suggests that while a fraction of star-forming galaxies may be quenched, the remaining galaxies undergo a boost of star formation. We discuss, in Sect. \ref{discussion}, the possible scenarii leading to this observation.\\
In the inner region of the tSZ signal where the Compton-$y$ is highest ($y>1.1\times10^{-5}$), we observe another change of regime in the weighted SFR (Fig. \ref{FigSZ_SFR_medians}, right panel). At Compton parameter $y_\mathrm{int} \approx 2. \times 10^{-5}$ ($\log y_\mathrm{int} \approx -4.7$) the weighted SFR of transitioning galaxies decreases slightly, in parallel we observed that the passive galaxies reach the same weighted SFR values. Since the intrinsic SFR is considerably lower for the passive galaxies, this indicates that their number density is considerably larger. Moreover above $y_\mathrm{int}$, we note that the SFR of the transitioning galaxies increases once again, the flat weighted SFR suggesting that the number density of transitioning decreases. 
At the same $y_{int}$ value, the weighted SFR of the star-forming galaxies does not change with $y$. Given that the SFR of these star-forming galaxies increases (Fig. \ref{FigSZ_SFR_medians}, left panel), this zero-slope relation is compensated by a decrease in the number density of star-forming galaxies.
This suggests that the drop seen in the weighted SFR is a combination of a strong decrease in the number of galaxies to compensate for the correlation between SFR and $y$. To complete these interpretations, we display, in Fig. \ref{Fig_nden_SFR_medians} (left panel), the relative fraction of passive, transitioning, and star-forming galaxies as function of the tSZ signal. In the left panel, we observed that galaxy SFR is strongly affected by the hot gas environment, with a dominant fraction of star-forming galaxies laying outside the outer contours of the tSZ signal ($y<2.2\times10^{-6}$) and passive galaxies dominating within the inner tSZ regions ($y > 1.1\times10^{-5}$). The intermediate zone between the outer tSZ contours and the innermost contours ($2.2\times10^{-6}<y<1.1\times10^{-5}$), marks the transition from a fraction of galaxies dominated by star-forming to a fraction dominated by passive galaxies.

\subsection{SFR--X-ray relation} \label{SFR-x}

By contrast with tSZ that traces extended region of hot gas around the SSC, the X-ray emission being proportional to the electron density squared traces the cluster cores of the SSC. Indeed as seen in Fig. \ref{FigFull_map}, the inner contours of the tSZ signal at $5\sigma$ level ($y_\mathrm{inner} \approx 1.1\times10^{-5}$) correspond to the outer contours of the X-ray emission at $1\sigma$ level (count rate (CR) such that CR$_\mathrm{inner}\approx$ 2000 counts s$^{-1}\ \mathrm{arcmin}^{-2}$). 
The inner X-ray contours at 5$\sigma$ level, defined by a count rate of CR$_\mathrm{outer}=10^{4}$ counts s$^{-1}\ \mathrm{arcmin}^{-2}$, are situated well within the core where Compton$-y$ values are the highest. 

\begin{figure*}
   \centering
   \includegraphics[width=0.45\textwidth]{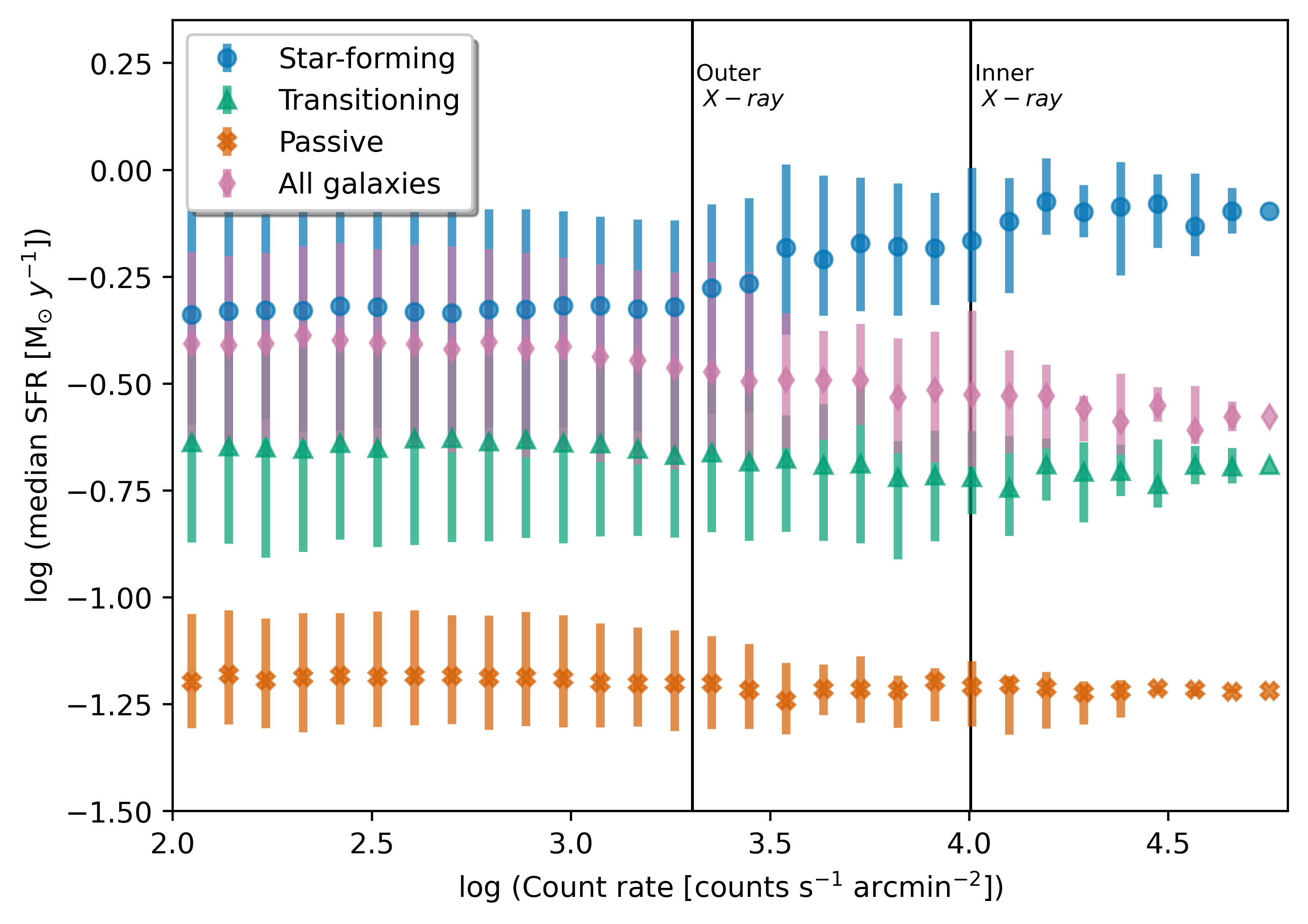}
   \includegraphics[width=0.45\textwidth]{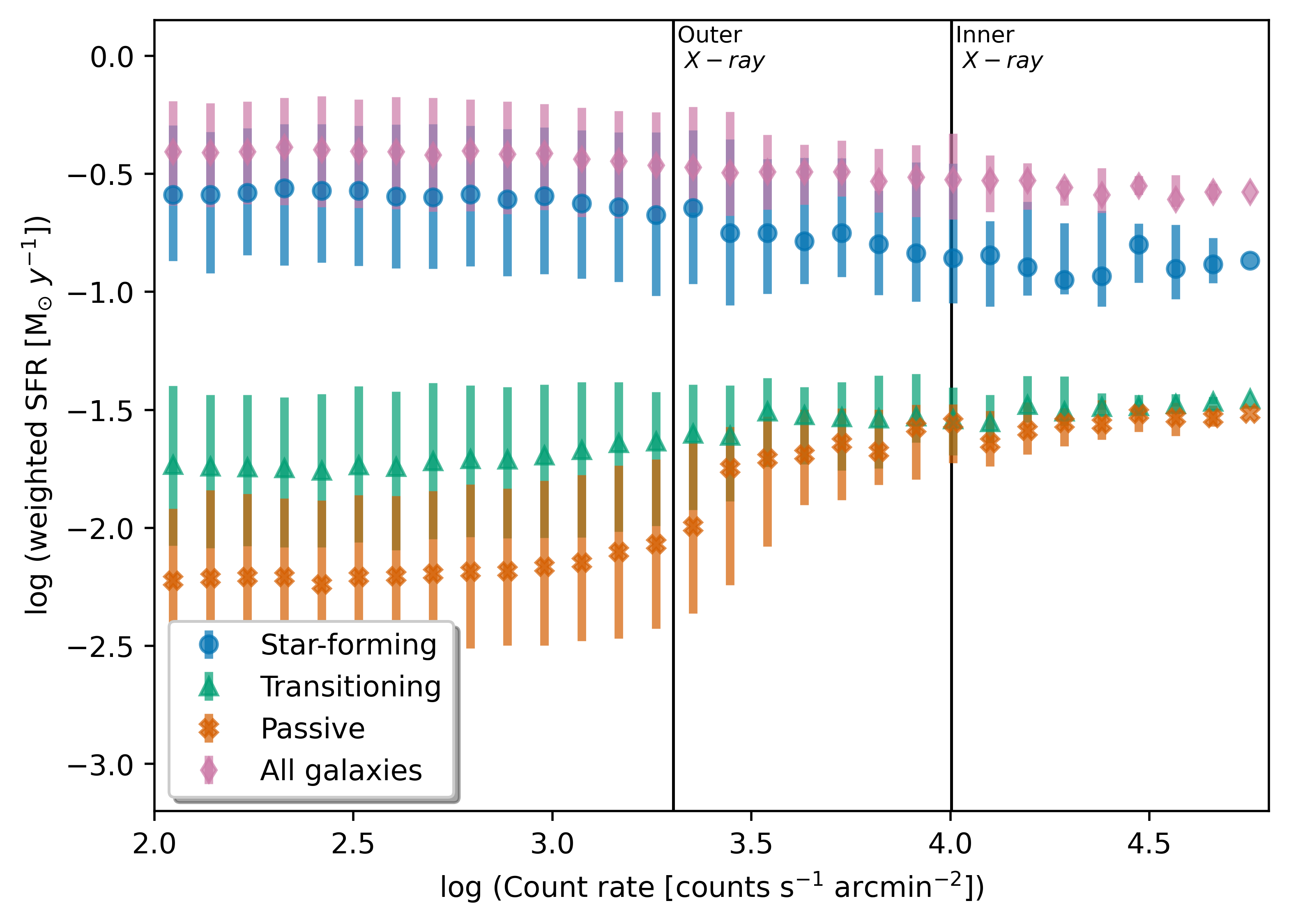}
   
   \caption{\textit{Left panel:} Median SFR values for all three populations (star forming, transitioning and passive) as a function of the ROSAT X-ray count rates. \textit{Right panel:} Distribution of the median SFR weighted by the fraction of galaxy-density of a given population w.r.t. total as a function of the ROSAT X-ray count rates. In both panels, the vertical lines correspond to the X-ray outer (CR$_\mathrm{outer}\approx 2000$ counts s$^{-1}\ \mathrm{arcmin}^{-2}$) and inner (CR$_\mathrm{outer}\approx 10^4$ counts s$^{-1}\ \mathrm{arcmin}^{-2}$) contours (see also text).}
    \label{FigX-ray_SFR_medians}%
\end{figure*}

We take advantage from the fact that X-ray emission traces essentially the cluster cores in the very central regions of the SSC to investigate the relation between star-formation activity and hot ionised gas. Considering the zoom-in region within $7 \times 6$ square degrees around the SSC, we repeat the analysis in Sect. \ref{SFR-y} and show in Fig. \ref{FigX-ray_SFR_medians} (left panel) the median SFR of all the galaxies as well as for each population (star forming, transitioning and passive) as a function of the X-ray count rates.\\
In the regime dominated by the background (CR<CR$_\mathrm{outer}$), the median SFR remains flat independent of the X-ray emission as observed in the SFR-$y$ relation (left panel in Fig. \ref{FigSZ_SFR_medians}). From the X-ray outer contours inwards, the general trend for all galaxies is a decreasing median SFR. The latter remains unchanged with increasing X-ray count rates all the way to the highest values, both for passive and transitioning galaxies. The star-forming galaxies, however, exhibit an increase in their median SFR which flattens out a first time, in the region CR$_\mathrm{outer}<$CR$<$CR$_\mathrm{inner}$, and second time in the very core region defined by CR$>$CR$_\mathrm{inner}$.

To account for the impact of galaxy-number densities on the median SFR, we computed the SFR weighted by the fraction of each galaxy population with respect to the total galaxy density (right panel in Fig. \ref{FigX-ray_SFR_medians}, similarly to Fig. 7). For all galaxy types, the weighted median SFR remains mostly unchanged (albeit a slight increase for the passive galaxies) for count rates below the outer X-ray contour. At larger count rates, the weighted median SFR of passive and transitioning galaxies increases with increasing X-ray emission. This is expected, since high count rates trace dense environments, within clusters, in which the number density of galaxies greatly increases. While this increase is very mild for the transitioning galaxies, the passive galaxies experience a rapid increase of the weighted median SFR up to the inner contour, after which it flattens out and reaches the same value as that of the transitioning galaxies. This behaviour is similar to the trend observed in the tSZ analysis where passive and transitioning galaxies had comparable value above $y_\mathrm{int}$. By contrast, the weighted median SFR of the star-forming galaxies gradually decreases with the X-ray count rate above CR$_\mathrm{outer}$. These observations may hint to the fact that, in the regions probed by the X-ray emission denser than those associated with the tSZ signal, most of the star-forming galaxies turn to passive (see Fig. \ref{Fig_nden_SFR_medians}, left panel). Hence, this would suggest that the fast quenching channel dominates in the environment traced by X-rays.

\subsection{Star formation activity in the SSC environment} \label{sec:SFdistrib}

\begin{figure*}
   \centering
   \includegraphics[width=0.45\textwidth]{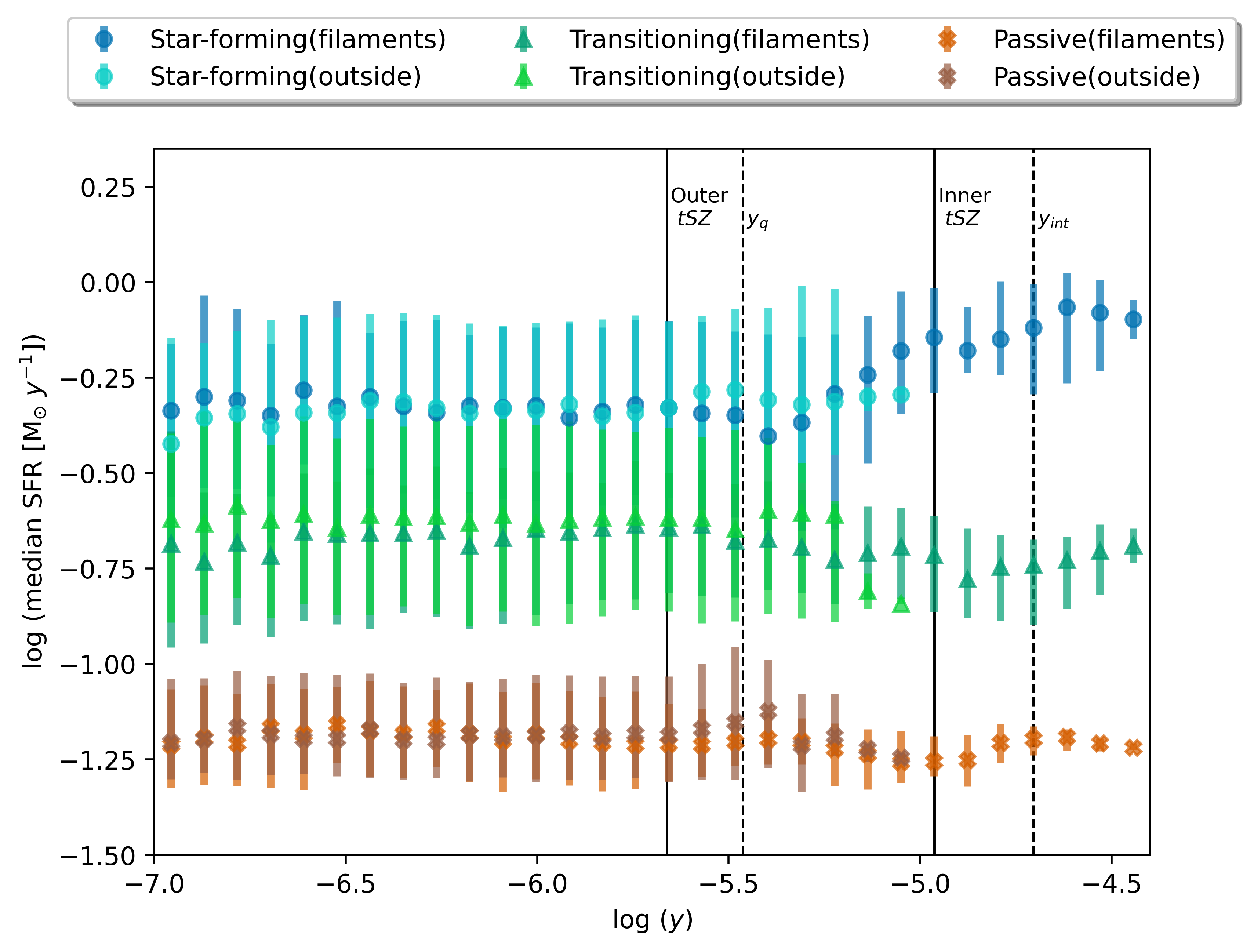}
   \includegraphics[width=0.45\textwidth]{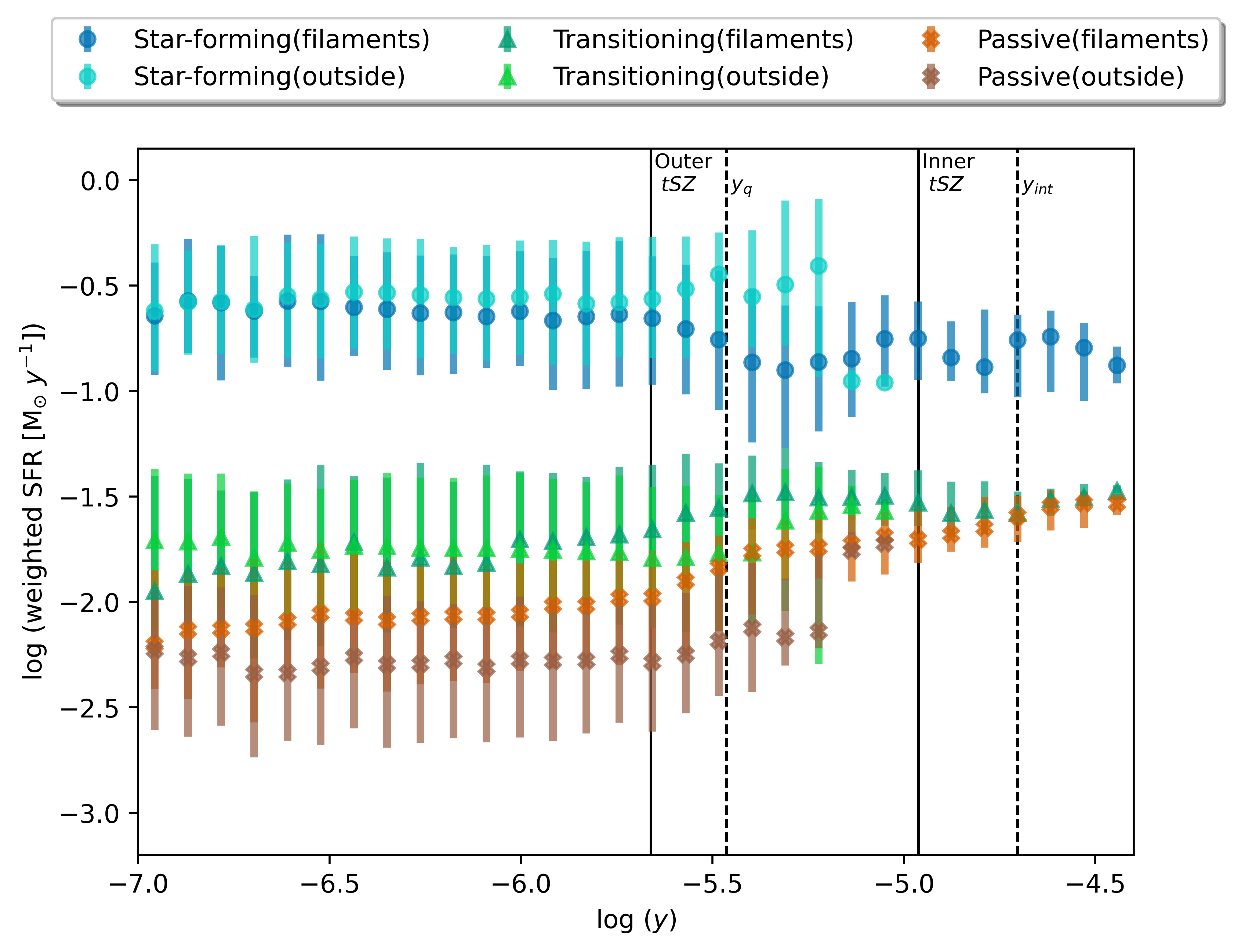}
   \caption{Same as Fig. \ref{FigSZ_SFR_medians}, but considering separately the regions within filaments (defined by the projection of the 3D T-REx filaments) and those outside the filaments.  
   }
              \label{FigSZ_SFR_filaments}%
    \end{figure*}

With the same relations computed in Sect. \ref{SFR-y}, namely SFR vs $y$ and galaxy-density weighted SFR vs $y$ to account for the variation of galaxy density of different types, we investigate the impact on the SFR-$y$ relation of environmental effects within and outside filaments. To do this, we compare the SFR (and weighted SFR) relations within and outside the area defined by the projected 3D filaments detected in Sect. \ref{fil-dect}. It is worth noting that the regions containing the highest Compton$-y$ values are all associated with the clusters and the connecting filaments in the very core region of SSC, i.e. inside filaments. This explains that the data-points outside filaments, in Fig. \ref{FigSZ_SFR_filaments}, do not extend to the highest $y$ values. Also throught this section, we focus solely on the tSZ signal given that all the regions defined by the outer X-ray contours are imbeded in the densest/core areas of the filaments.

For passive galaxies, we see that the SFR-$y$ relation (Fig. \ref{FigSZ_SFR_filaments}, left panel, yellow and red crosses) is the same inside and outside filaments for the whole range of $y$ values, with a small deviation (decrease of SFR) from a zero-slope relation at $y_q$.  \\
In the case of the transitioning galaxies, the SFR-$y$ relation (Fig. \ref{FigSZ_SFR_filaments}, left panel, green and magenta triangles) is also the same inside and outside filaments up to $y_q$, where the SFR outside filaments becomes slightly larger than inside filaments. In the region defined by the tSZ inner contours ($y>y_{inner}=1.1\times 10^{-5}$), the SFR-$y$ relation has a zero slope but the overall median SFR values are smaller indicating that the the star-formation activity has decreased in the high-Compton parameter region, due to quenching.\\
The star-forming galaxies inside (blue dots) and outside (pink dots) filaments exhibit the same median SFR values, independently of $y$ (Fig. \ref{FigSZ_SFR_filaments}, left panel). It is as expected larger than the two other populations. Beyond $y_q \approx 3.4 \times 10^{-6}$, the SFR-$y$ relation inside and outside differs with an increase
of SFR inside filaments, while it remains flat outside. In the regions delimited by the inner tSZ contours and where the Compton parameter values are the largest, we witness at $y > y_\mathrm{inner}=1.1\times10^{-5}$ a steepening of the SFR vs $y$ slope and the median SFR reaches larger values indicating a significant star formation activity. 

\bigskip
A complementary view is provided by the analysis of the SFR weighted by the number density of galaxies for each population (Fig. \ref{FigSZ_SFR_filaments}, right panel). 
We can see in particular that the star-forming galaxies inside (blue dots) and outside (pink dots) filaments share the same slopes and median SFR values up to $y_\mathrm{outer}\approx 2.2\times 10^{-6}$ where they start departing from each other. Beyond this value, the weighted SFR inside filaments (blue dots) gradually decreases with $y$ due to the decrease in the number density while the weighted star formation outside filaments (pink dots) increases.
In the very core region of SSC $y>y_\mathrm{inner}=1.1\times 10^{-5}$, the number-density weighted SFR decreases slightly
due to the combined effect of reduced star formation and number of galaxies. \\
Focusing on the weighted SFR-$y$ relation for transitioning galaxies displayed in Fig. \ref{FigSZ_SFR_filaments} (right panel), we observe that the SFR of galaxies outside (green triangles) filaments do not vary with $y$ while for galaxies inside (magenta triangles) filaments there is a gradual and slow increase up to $y_q$. At this Compton parameter value, the weighted SFR outside filaments gradually increases with $y$. On the other hand within filaments, the SFR slightly decreases between $y_q < y < y_\mathrm{int}$, after which it starts increasing again. \\
In parallel, we see that the weighted SFR of passive galaxies follow a similar evolution with $y$, albeit larger median SFR inside the filaments (yellow crosses) than outside (red crosses). Indeed, we observe a slight increase in the SFR for Compton parameter values $y>2.2\times 10^{-6}$ (within the outer tSZ contour). At $y_\mathrm{int}\approx 1.62 \times 10^{-5}$, the weighted median SFR in passive galaxies reaches that of the transitioning galaxies. This indicates that the number density of passive galaxies is much larger than of transitioning galaxies, compensating for the SFR of the former which is much lower (left panel in Fig. \ref{FigSZ_SFR_filaments}). This larger number of passive galaxies is likely the result of an increased number of transitioning galaxies turning passive or star-forming turning directly passive. These interpretations are in line with the variation of the relative fraction of passive, transitioning, and star-forming galaxies as a function of the tSZ signal inside and outside filaments shown in Fig. \ref{Fig_nden_SFR_medians} (right panel).

%
%

\section{Discussion}\label{discussion}
\begin{itemize}
    \item[$\bullet$] We derived the large scale structure of the SSC using a state-of-the-art web-finder algorithm, T-REx, applied both on 2D and 3D galaxy distributions. Our results agree very well with those of \cite{quintana20}. 
    Not only we confirm with a blind detection technique the main structure of the filamentary pattern; but we also identify reliable filaments connecting the SSC central region to the clusters surrounding it. Additionally, we detect the filaments within the large complexes of galaxies in the foreground and background of SSC. T-REx outputs an associated probability to the detected filaments which allows us to estimate their reliability. We found that low-reliability filaments connect the SSC region to the two, foreground and background, galaxy complexes.   
    We have compared our results with those of \cite{higuchi2020} on the DM distribution in 23 square degrees around the SSC from a weak lensing (WL) analysis based on the data from \cite{merluzzi2015} and \cite{haines2018}. \cite{higuchi2020} show internal (within Shapley core) and external filaments in the reconstructed projected mass distribution in the r and g photometric bands. The overall agreement with our results, based on T-REx, is very good in the core region of Shapley. In the outskirt notably, we clearly detect quantitatively with T-REx applied on the 3D galaxy distribution, the filament that relates A3558 and A3559. This structure was also seen as a stream of galaxies by \cite{haines2018}. We find a good overlap between this stream and a 2D filament detected with T-REx. Comparing with the 3D filament, the stream in \cite{haines2018} connects the two clusters more towards the eastern side, while T-REx 3D filament is directly to the north. However, it is worth noting that the T-REx filaments quoted here are identified as reliable when their probability is greater than 0.5. Hence, the difference between our two results may be due to the presence of a lower probability filament traced in \cite{haines2018}. In addition, it was hinted by the WL analysis performed in the r band (see Fig. 4 in \citealp{higuchi2020}) but given the low density contrast no definitive conclusion was made. In addition to these filaments, we also detect quantitatively with T-REx the two filaments relating, on the one hand, Shapley core with A3554 and, on the other hand, Shapley core with A3560. 
\bigskip
    \item[$\bullet$] The SSC area is a large and complex concentration of galaxies. Its analysis using the T-REx filament-finder technique permits us to investigate several aspects of the connection of clusters to their large-scale cosmic web environment. For instance, the number of filaments connected to clusters, i.e. the connectivity, has been the subject of different studies in simulation \citep{calvo2010,codis2018,gouin2021,Kraljic2022,malavasi2023,galarraga2023,galarragaTNG2023} and observations \citep{DarraghFord2019,Sarron2019,malavasi2020,maret-2020,maret2021}. In particular, \cite{gouin2021} have used T-REx technique on simulations and shown that the connectivity relates to mass assembly in clusters and hence their mass and dynamical state; and that it impacts observational properties such as their shapes, etc. \citep[see also][]{gouin2022}. Our analysis of the Shapley field identified the filaments connecting the galaxy clusters in the central part of the supercluster. It also showed that Shapley's central region is connected to five reliable filaments. Four of them, shown in Figs. \ref{FigFilament_3D} and \ref{FigPopulations}, were already hinted in previous studies (see Sect. \ref{sec:struc}), the fifth is a filament on the western side of the SSC. This connectivity is coherent with that of supercluster A2142, estimated to 6-7 filaments according to \cite{maret-2020}, or that of Corona Borealis estimated to seven \citep{maret2021}. As shown in the analysis of numerical simulations \citep{gouin2021} these high values of the connectivity are representative of massive systems still ongoing mass accretion, i.e. still forming, as suggested by \cite{chon2015} for the Shapley supercluster. \\
    
    Filaments of the cosmic web not all have the same nature \citep[e.g.,][]{job2023}. In addition, \cite{galarraga2020} showed that two extreme populations of filaments trace different environments and have different physical and thermodynamical properties \citep{galarraga-gas2021,toni2021,galarragaTNG2023}.  
    Short filaments ($\approx 10$~Mpc) trace dense environments relating massive galaxy clusters and are filled with hot gas. In the very central region of Shapley, we clearly detect this type of filaments, connecting A3558 to A3562, and emitting in the X-rays \citep{shapleyX1999,merluzzi2016}. This very central region with the highest density of matter filled with hot gas that also shows the largest tSZ signal ($y>10^{-5}$) seen in the \textit{Planck} data \citep{plck-shapley2014}.  
\bigskip
    \item[$\bullet$] The impact of the filaments detected in the cosmic web on galaxy properties, in particular star-formation, has been shown by many analyses both in simulations and in observations \citep[e.g.,][]{poudel2017,laigle2018,bonjean2020,malavasi2022,bulichi2023,Herzog2023,Ramsoy2021,galarraga2023}. It was also observed in the outskirts of clusters out to several virial radii, again both in simulations and observations \cite[e.g.,][]{salerno2019,Sarron2019,gouin2020,kuchner2020,sperone2021,cornwell2023}. One of the major impacts is the pre-processing process by which star formation is quenched before galaxies fall in clusters, with the observation in the cluster outskirts at large virial radii that the density of star-forming galaxies does not reach the level of the field star-forming galaxies.     
    In our study when considering all the galaxies in the SSC field, there is a much larger number of star-forming than passive galaxies and we also find that the outer regions of the SSC have similar fractions of star-forming galaxies than the field (see Fig. \ref{Fig:d2ms_histogram}). However within regions defined by the detected filaments, the number of passive and star-forming galaxies is comparable. This indicates that the population of galaxies within filaments is quenched, suggesting that galaxies are pre-processed in the filaments and the groups therein \cite[in agreement with e.g.,][]{detachmnt2019,lopes2023}. \\
    In parallel to the decrease of the weighted SFRs of star-forming galaxies (Fig. \ref{FigSZ_SFR_medians}), our results also show that both weighted SFRs of passive and transitioning galaxies increase when getting closer to the inner regions of the SSC, until a $y$ threshold beyond which passive and transitioning galaxies reach the same weighted median SFR. This confirms the impact of the dense environment in the SSC core in quenching the star formation. It may also indicate that the two quenching channels, fast and slow, discussed in \cite{moutard2018} are at play in the Shapley supercluster quenching the star-forming galaxies and therefore increasing in parallel the number of transitioning and passive galaxies. A similar scenario may also be at play in the Virgo superclusters, where \cite{chung2021} showed that about half the galaxies in the VirgoIII filament are transitional dwarf galaxies, transforming into quiescent dwarf early-type galaxies. By nature of the Bremsstrahlung emission, the regions of the SSC probed by the X-ray emission correspond to denser environments than those traced by the tSZ signal. In these regions, we observe a milder evolution of the weighted median SFR with X-ray count rates both for star-forming and transitioning galaxies while it increases more sharply for passive galaxies. This may hint to the fact that most of the star-forming galaxies turn to passive via the fast quenching channel proposed in \cite[e.g.,][]{moutard2018}.

\bigskip
    \item[$\bullet$] The environment of galaxies (in other words the cosmic web structures, in particular clusters and filaments) is known to impact galaxies by triggering or quenching star formation via many interlinked and different physical processes (galaxy encounters or mergers, ram-pressure stripping, gas evaporation, etc.) \cite[e.g.,][and references therein]{2022Univ....8..554A}. The local density of galaxies has long been identified as a main driver of the quenching/star-formation processes \citep[][and references therein]{2022A&A...668A..69E}. Consequently, galaxy environments are traditionally defined by their local density and can also be defined by the luminosity density \citep[e.g.,][]{liivamagi2012,teet2020}. However, the gas properties of the environments in which galaxies are located, in particular the temperature, are also known to impact galaxies \cite[e.g.,][]{dave2015}. \\
    In our study, we define the environment in which galaxies evolve combining both density and gas temperature. This is conveniently captured by the Compton parameter, which measure the integrated gas pressure. We hence use the $y$ parameter, which measures the amplitude of the tSZ signal, as a proxy of the zone of influence of the cosmic web matter content both in clusters and filaments. Moreover rather than a commonly used distance-centric definition, we define the zone of influence of the environment via the contours of the $y$ parameter amplitude. \\ 
    Therefore, we analysed the impact of the environment on the star-formation activity as a function of the Compton parameter.     
    We observed that, beyond a given $y$ value equivalent to $2\sigma$ (where $\sigma$ defines the background tSZ signal), the median SFR decreases with increasing tSZ signal in particular in the very central core of the Shapley supercluster. In agreement with \cite{bonjean2020}, we also find that the regions of SSC which are filled with hot gas, traced by the tSZ signal, contain mostly low star-forming or transitioning galaxies. In \cite{bonjean2020}, the authors performed a study of the distribution of quiescent galaxies w.r.t. the filament spines that they compared to observed tSZ signal in the filaments from \cite{hidekiSZ2020}. Their result suggested that the hot gas around filaments has a non-negligible role in quenching star formation. The observed absence of blue late-type spirals together with the shaping of the galaxy luminosity functions, by \cite{mercurio2006}, also hinted towards the role of mechanisms related to the gas such as galaxy harassment or ram-pressure stripping in reducing or stopping star formation. In our study, we find a similar trend (see Sects. \ref{SFR-y} and \ref{sec:SFdistrib}). 
    Certainly at the very highest $y$ values, the fraction of passive galaxies in the SSC dominates over both star-forming and transitioning populations. Moreover for the  star-forming galaxies, we identify a threshold Compton parameter value, $y_{q}$, that coincides with an increase in the SFR suggesting that while a large fraction of star-forming galaxies may be quenched, the remaining galaxies undergo a boost of star formation. 
    
    As suggested by \cite{singh2020}, this could be the result of condensation of intra-filamentary gas into the filament galaxies fuelling star formation; or it could be the result of interactions between filament galaxies due to increased number density of galaxies near the filament spines. In this case, mergers convert star forming spirals into passive ellipticals, while interactions stir the ISM to enhance star formation. So at the same time star-forming galaxies are less numerous but star-formation is very efficient. Merger-induced star formation has been observed in other cases \cite[e.g.,][]{johnston-merger2008}. 
    In our analysis, the observed increase of the median SFR above $y_q=3.4\times 10^{-6}$ takes place mainly in the core region of Shapley and its connected filaments, recognised as undergoing gravitational collapse \cite[][]{reisenegger2000,Ragone2006,chon2015}, which would indeed suggest merger-driven star formation. The associated decrease of weighted median SFR for the star-forming galaxies, observed at $y_q$, could thus be explained by the competition between on the one hand the effect of merging/interactions, increasing the SFR, and on the other hand the higher gas pressure inside the core region of Shapley, stripping off the cold gas from galaxies and thus preventing star formation. This could be comparable to the findings of \cite{castignani2022}, in the case of Virgo, where they conclude that gas is depleted in filaments before the quenching processes, depending mostly on the local density. In the Shapley supercluster field and particularly at the connection and within the very core of SSC, the filamentary network is likely made of prominent, short filaments tracing the densest surroundings \cite[e.g.,][for a definition of short filaments]{galarraga2020,job2023} and expected to be filled with gas at temperature $> 10^6$~K or larger \citep{gheller2015,gheller2016,Gheller2019,martizzi-gas2019,galarraga-gas2021,toni2021,gouin2022,Vurm2023}. An analysis of the thermal properties of gas in prominent filaments in hydrodynamical numerical simulations by \cite{zhu2022} shows that about half the gas, located in relatively highly overdense regions, is hotter than $10^6$~K. \cite{zhu2022} propose that this hot gas could suppress the supply of cold gas to halos before they enter clusters.  This physical process could hence explain the quenching of the star formation in the galaxies located in the Shapley supercluster's environment, triggering the relation we observed between median SFR and Compton-$y$ parameter. Hydrodynamical simulations also suggest that gas accretion of galaxies decreases closer to filaments, due to the vorticity at the edges of filaments \citep{Song2021}. This could act as an additional source of quenching.   

\end{itemize}

%
\section{Conclusions}\label{conclusion}

The present analysis focuses on the study of the Shapley supercluster in its large-scale environment with two main goals. First, we aimed at identifying quantitatively the large-scale filamentary network to which the Shapley multiple-cluster system is connected. Then, we aimed at investigating how the environment defined by the SSC, together with the filaments attached to it, modifies the star-formation activity of the galaxies located inside and outside Shapley and its filamentary network.

We investigated the structure of the whole Shapley supercluster system based on the 2D photometric and 3D spectroscopic galaxy distributions within $15 \times 15$ square degrees. By applying a dedicated filament-finder method based on graphs, T-REx, to the 2D and 3D galaxies, we traced in a quantitative manner the filamentary network in the SSC and around it. Our filament finder quantifies the reliability of the detected filaments by computing an associated probability to their detection. We detected all the filaments found or hinted in previous studies confirming their reliability. We also detected new filaments which were not previously observed, connecting the SSC central region to the clusters surrounding it. We found that the Shapley core region connecting A3558 to A3562 has a connectivity of five as expected for a system ongoing collapse. In addition, we detected the filamentary network of large complexes of galaxies in the foreground and background of SSC.

Focusing on the star-formation rate in star-forming, transitioning and passive galaxies in a zoom-in region of $7\times6$ square degrees centred on the SSC core, we exhibited the impact of the SSC environment. Within the areas defined by the detected filaments, we find that the
number of passive and star-forming galaxies is comparable while it is the opposite in the outer regions of SSC indicating that the population of galaxies within filaments is pre-processed and quenched. 

Both the density and the gas properties, in particular temperature, of the environments affect star-formation activity. The amplitude of the tSZ effect, i.e. Compton $y$ parameter, measures the integrated gas pressure that is a combination of density and temperature. Therefore, we defined environment of the SSC and its filamentary network based on the amplitude of the tSZ effect, as a proxy of the zone of influence of both the clusters and the filaments. We then analysed the impact of the environment on the
star-formation rate of galaxy population as a function of the Compton $y$ parameter reconstructed from the \textit{Planck} data. We found that the zone of influence of the supercluster on the galaxy properties extends beyond the core region of Shapley. As expected, we showed that the star formation rate decreases significantly with increasing tSZ signal in particular in the very central core of SSC which contains mostly passive or transitioning galaxies. However, we also identified a threshold Compton $y$ value where the median SFR of star-forming galaxies increases. This can be explained by the fact that while a large fraction of star-forming galaxies are quenched, potentially via both slow and fast channels, the remaining ones undergo a boost of star formation probably induced by the interactions and mergers in the SSC core region. All these results were confirmed by our analysis of the star-formation rate as a function of the X-ray emission in the SSC field, from ROSAT, with the main difference being that galaxies seem to be quenched mainly via the fast channel, that is, directly from star forming to passive. This could be explained by the denser regions probed by X-rays not allowing for a slow, gradual transition to passivity. 

\medskip
The analysis of the relation between the median star-formation rate of galaxy populations and the integrated pressure of the gas, contained in the clusters and the filaments connected to them, indicates that both the distribution of galaxy types inside and outside the SSC filamentary network and the star-formation activity are the result of a competition between the effects of merging/interactions, increasing the SFR, and the higher gas pressure inside the SSC structure, preventing star formation. In this context, the use of the tSZ signal to define the zone of influence of the cosmic-web structures is a novel way of identifying the environment of galaxies. The combination and comparison of data from existing and future tSZ surveys, such as those of the Atacama Cosmology Telescope\footnote{\url{https://act.princeton.edu/}}, the Simons Observatory\footnote{\url{https://simonsobservatory.org/}}, or the South Pole Telescope\footnote{\url{https://pole.uchicago.edu/public/Home.html}}, and X-ray surveys such as \textit{eROSITA}\footnote{\url{https://erosita.mpe.mpg.de/}}, with those of the large galaxy surveys, e.g. Euclid\footnote{\url{https://www.euclid-ec.org/}}, Dark Energy Spectroscopic Instrument\footnote{\url{https://www.desi.lbl.gov/}}, etc., should allow us to address in a systematic and statistical manner the complex topic of galaxy evolution across all cosmic web environments.

\begin{acknowledgements}
This research was supported by funding for the ByoPiC project from the European Research Council (ERC) under the European Union’s Horizon 2020 research and innovation program grant number ERC-2015-AdG 695561.
T.B. acknowledges funding from the French government under management of Agence Nationale de la Recherche as part of the ``Investissements d’avenir'' program, reference ANR-19-P3IA-0001 (PRAIRIE 3IA Institute). The authors thank D. Eckert and  M. Douspis for comments together with the members of the ByoPiC team\footnote{\url{https://byopic.eu/team}} for useful discussions. They also thank an anonymous referee for their comments and suggestions.
\end{acknowledgements}

\bibliographystyle{aa} 
\bibliography{references} 

\end{document}